\documentclass[]{emulateapj}

\usepackage{times}
\usepackage{amsmath,amssymb,amsfonts}
\usepackage{graphicx}
\usepackage{color}
\usepackage[varg]{txfonts}
\usepackage{enumerate}
\usepackage{enumitem}
\usepackage{placeins}
\newcommand{\mlr}{${\cal M}/{\cal L}$}
\newcommand{\smlr}{${\cal M}/{\cal L}_\star$}

\newcommand{\cena}{Centaurus\,A}

\def\ltsima{$\; \buildrel < \over \sim \;$}
\def\simlt{\lower.5ex\hbox{\ltsima}}
\def\gtsima{$\; \buildrel > \over \sim \;$}
\def\simgt{\lower.5ex\hbox{\gtsima}}
\def\hide#1{}

\begin{document}

\title{The Lost Dwarfs of Centaurus A and the Formation of its Dark Globular Clusters}

\author{Mia Sauda~Bovill$^{1,2}$, Thomas H.~Puzia$^{2}$, Massimo Ricotti$^3$, Matthew A.~Taylor$^{2,4}$}
\affil{1 - Space Telescope Science Institute, 3700 San Martin Drive, Baltimore, MD 21218, USA}
\affil{2 - Institute of Astrophysics, Pontificia Universidad Cat\'olica de Chile, Av.~Vicu\~na Mackenna 4860, 7820436 Macul, Santiago, Chile}
\affil{3 - Department of Astronomy, University of Maryland, College Park, MD 20740, USA}
\affil{4 - European Southern Observatory, Alonso de Corboda 3107, Vitacura, Santiago, Chile}
\email{bovill@stsci.edu}



\label{firstpage}

\begin{abstract}
We present theoretical constraints for the formation of the newly discovered dark star clusters (DSCs) with high mass-to-light (${\cal M}/{\cal L}$) ratios, from Taylor et al (2015). These compact stellar systems photometrically resemble globular clusters (GCs) but have dynamical ${\cal M}/{\cal L}$ ratios of $\sim\!10\!-\!100$, closer to the expectations for dwarf galaxies. The baryonic properties of the dark star clusters (DSCs) suggest their host dark matter halos likely virialized at high redshift with ${\cal M}\!>\!10^8\,M_\odot$. We use a new set of high-resolution N-body simulations of Centaurus A to determine if there is a set of $z\!=\!0$ subhalos whose properties are in line with these observations.~While we find such a set of subhalos, when we extrapolate the dark matter density profiles into the inner $20$ pc, no dark matter halo associated with Centaurus A in our simulations, at any redshift, can replicate the extremely high central mass densities of the DSCs. Among the most likely options for explaining $10^5\!-\!10^7\,M_\odot$ within 10 pc diameter subhalos is the presence of a central massive black hole. We, therefore, propose that the DSCs are remnant cusps of stellar systems surrounding the central black holes of dwarf galaxies which have been almost completely destroyed by interactions with Centaurus A.
\end{abstract}


\section{Introduction}
\label{SEC.intro}

The formation of compact stellar systems (CSSs), such as globular clusters (GCs), ultra-compact dwarfs (UCDs), and dwarf-globular transition objects (DGTOs) is a major contemporary problem for modern numerical simulations \citep[e.g.][and references therein]{ren13, ren15}. Recent studies have revealed surprisingly complex properties, such as central black holes (BHs) and extended star-formation histories \citep[e.g.][]{set14, nor15}.

Recently, \cite{tay15} (hereafter T15) derived dynamical masses for 112 of the brightest CSSs in the nearby giant elliptical galaxy NGC\,5128, \citep[\cena; $D\!=\!3.8\!\pm\!0.1$\,Mpc,][]{har10}.~In addition to the expected GCs and UCDs, they found evidence for so-called dark star clusters (DSCs) which exhibit disproportionately high mass-to-light (${\cal M}/{\cal L}$) ratios but have typical Local Group GC luminosities and structural properties.~In comparison to UCDs, which show a scaling relation as ${\cal M}_{\rm dyn}/{\cal L}\!\propto\!{\cal M}_{\rm dyn}^{0.33\pm0.04}$, the DSC sequence exhibits a significantly steeper scaling, i.e.~$\propto\!{\cal M}_{\rm dyn}^{0.79\pm0.04}$.~No peculiarities in the spatial distribution of these DSCs compared to UCDs or the normal GC population were found.~The possible origins of DSCs are presently unclear.~Excluding observational biases \citep[see][for a detailed discussion]{tay15}, it is thought that a combination of several physical mechanisms could explain the high \mlr\ ratios of DSCs: {\it i)} systemic rotation, {\it ii)} significant dark matter (DM) content, and/or {\it iii)} central massive BHs, however, the validity of these explanations has not been tested in numerical simulations.

First, we look at rotation in globular clusters. Some Milky Way GCs show slow rotation with amplitudes of several km s$^{-1}$ \citep[e.g.][]{bia13, kac14}, which are at least one order of magnitude slower than what would be required to fully explain the \mlr\ ratios of DSCs.~In fact, the required rotational velocities would destabilize the DSCs, and likely unbind the systems (T15).~With that in mind, DSC rotation cannot be entirely ruled out, but rotation and tri-axiality cannot entirely account for their elevated \mlr\ ratios.

We next discuss the possible influence of dark matter on globular cluster formation and evolution. It has been proposed that Milky Way GCs may have formed within DM halos \citep{pee84, bro02}. If so, they must have lost the majority of the dark matter via dynamical processes throughout their evolution, since today there is no evidence for significant DM content in GCs \citep[e.g.][]{Moore:96, Baumgardtetal:09, lan10, Conroyetal:11, fen12, iba13, hur15}. For a discussion of MOND-based explanations, see \cite{dar14}, however, MOND predictions fall orders of magnitude below what is required to explain the derive DSC dynamics (Smith \& Candlish, private communication).

Finally, we discuss the evidence for and implications of central IMBHs in GCs. Observational limitations and biases \citep[e.g.][]{vdb10, bia15} have prevented the confirmation of central intermediate-mass BHs (IMBHs) in Local Group GCs \citep[e.g.][]{sun13, kam14, lue15, cse15, wro15}.~For example, the Milky Way GC M10 does not seem to possess the cuspy central surface density profile indicative of a central black hole, but the presence of an IMBH with a mass up to 0.75\% of the total GC mass, corresponding to ${\cal M}_\bullet\!\approx\!600\,M_\odot$, cannot be excluded \citep{umb13}.~In any case, BH signatures in more massive CSSs have been definitively detected \citep{geb05, vdm10, set14}. A central cluster of BHs would not necessarily produce the cuspy observational signature of an IMBH \citep{BanerjeeK:11}, however such a dark cluster would be unable to duplicate the extremely high central densities derived for the DSCs.

The satellite system of \cena\ within $\sim\!50\,{\rm kpc}$ of galactocentric distance is dominated by its rich known GC system, with over 600 confirmed members \citep{van81,hes84,hes86,har92,pen04,woo05,woo10,rej07,bea08}, many of which populate the high-luminosity (ie. high mass) end of the globular cluster luminosity function (GCLF). The total expected population may be as high as 2000 \citep{har84,har02,har10,har12}, but the distribution at large galacto-centric radii is relatively unconstrained. At these large radii, the known satellites of \cena\ are limited to a few dozen confirmed dwarf galaxies within $\sim$1.4\,Mpc of galactocentric distance \citep[e.g.][]{cot97,ban99,jer00,kar02,kar07,crn14,tul15}, but this population may not be complete due to the small sky coverage and shallow surface brightness limits of previous photometric surveys.

In this work, we present results from high-resolution $\Lambda$CDM numerical simulations of a \cena\ analog halo. Based on our results we put upper limits on the central DM densities in the putative DSC candidates, and, by inference, provide stringent direct estimates of potential central BH masses.~This paper is laid out as follows: Our high-resolution simulations of the \cena\ analog halo are described in \S\ref{SEC.simulations}, and the constraints provided by the known observed properties of the DSCs are given in \S\ref{SEC.constraints}. We highlight the relevant circular velocity and density profiles of the dark matter halos compared to the measured stellar velocity dispersions of the DSCs in \S\ref{SEC.DM}, and the possibilities for the formation and growth of a central massive black hole in these systems in \S\ref{SEC.seed}. Finally, observational tests for our model are provided in \S\ref{SEC.tests} and our summary and conclusions are presented in \S\ref{SEC.conclusions}.

\section{Simulations}
\label{SEC.simulations}
In this section, we describe a suite of high resolution cosmological simulations of an isolated Centaurus A analog from $z\!=\!150$ to $z\!=\!0$.~All initial conditions were generated with {\sc MUSIC} \citep{HahnA:11} and simulations were run with {\sc Gadget 2} \citep{spr05}.~Halo properties were analyzed using {\sc Amiga} \citep{Gilletal:04,KnollmanK:09} and merger trees generated with {\sc consistent\_trees} \citep{Behroozietal:13}. Code testing and preliminary simulations were performed on the {\sc Geryon} cluster at the Institute of Astrophysics at Pontificia Universidad Cat\'olica de Chile, and all subsequent high-resolution simulations were run on the University of Maryland HPCC {\sc Deepthought 2}.

We initially ran a 100 Mpc/h box with $N\!=\!256^3$, resolving \cena\ candidates at $z\!=\!0$ with $>1000$ particles.~All simulations are run from $z\!=\!150$ to $z\!=\!0$ with outputs every $10$ Myr for $z\!>\!6$ and every $100$~Myr for $z\!<\!6$.~Our \cena\ analog halo is chosen to have a total virial mass of $\sim\!10^{13} M_\odot$, corresponding to the upper range of virial masses measured for \cena\ \citep{van00,pen04, woo07, woo10, lok08, har14}. Our \cena\ analog is also roughly consistent with, if slightly more centrally concentrated than, measurements of $v_{max} \sim 100$km~s$^{-1}$ at $7-22$~kpc \citep{Israel:98} and $\sim3\times10^{11} M_\odot$ within $\sim50$~kpc \citep{Mathieuetal:96}.~To approximate the Local Volume, we use an isolation criterion of no halos with ${\cal M}_{\rm vir}\!>\!10^{12} M_\odot$ within $3$~Mpc/h of the \cena\ analog at $z\!=\!0$.~We do not account for the presence of M83, which is $\sim\!1$\,Mpc away from \cena; however, this should not bias our results significantly, as M83 is $\sim\!2\!-\!3\,R_{\rm vir}$ away from \cena.~As we focus on the inner virial halo of \cena\ ($<350$\,kpc) regions of Centaurus A, thus the presence (or lack) of an M83 analog will not change the properties and distribution of subhalos within our region of interest.~This assumption closely matches those adopted while running an isolated Milky Way analog without an M31 counterpart to study Galactic satellite populations \citep[e.g.][]{bul05, joh08,Springeletal:08,Diemandetal:07}.

Once we identify our \cena\ analog, we resimulate the Lagrangian region corresponding to $\sim\!4\,R_{\rm vir}$ ($\sim\!2.4$\,Mpc) at $z\!=\!0$ at higher resolution to produce a clean region of $r\!\simeq\!2\,R_{\rm vir}$ at $z\!=\!0$. We perform two resimulations, at $N_{\rm eff}\!=\!4096^3$ with $m_p\!=\!1.13\times10^6\,M_\odot$ and $\epsilon\!=\!500$~pc physical (run 100M\_4096) and at $N_{\rm eff}\!=\!8192^3$ with $m_p\!=\!1.4 \times 10^5\,M_\odot$ and $\epsilon\!=\!200$~pc physical (100M\_8192).~These resolutions were chosen to resolve halos with $M>10^8 M_\odot$ with $100$ and $1000$ particles respectively.~$N_{\rm eff}$ is the number of particles in the simulation if the entire $100$~Mpc/h box were simulated at the highest resolution and $\epsilon$ is the gravitational softening length.~Our two re-simulations resolve halos with masses above $1.13\!\times\!10^8\,M_\odot$ and $1.4\!\times\!10^7\,M_\odot$, respectively, with 100 particles.~Table~\ref{TAB.runs} provides a summary of these simulations and the properties of the respective \cena\ analogs. All halo parameters are those derived by {\sc Amiga} for the Centaurus A analog at $z=0$. $R_{vir}$ of a given halo depends on both $M_{vir}$ and $v_{max}$, but in a non-linear way.

\begin{table*}
	\centering
	\caption{Summary of simulation characteristics. Columns in order are (1) label for the simulation, (2) maximum effective resolution, (3) minimum dark matter particle mass, (4) physical softening length in pc/h, (5) Number of particles in the \cena\ halo, (6) Virial mass of the \cena\ halo in units of $10^{13} M_\odot$, (2) maximum circular velocity of the \cena\ halo in km s$^{-1}$, and (7) Virial radius of the \cena\ halo in kpc/h.}
	\begin{tabular}{c c c c c c c c c} \hline
	\hline
	Run & $N_{\rm eff}$ & $m_p$ & $\epsilon$ & $N_{\rm part}$ & $M_{\rm \cena\ }$ & $v_{max}^{\rm \cena\ }$ & $R_{\rm vir}$ \\ 
	 &  & ($M_\odot$) & (${\rm pc}$) & & ($10^{13} M_\odot$) & (km~s$^{-1}$) & (kpc~h$^{-1}$) \\
	\hline
	100M\_4096 &  $4096^3$ & $1.2\times10^6$ & $500$ & 10334710 & 1.16444 & 437.33 & 368.60 \\
	100M\_8192 & $8192^3$ & $1.4\times10^5$ & $200$ & 83263765 & 1.17269 & 437.78 & 369.46  \\
	\hline 
	\hline
	\end{tabular} 
\label{TAB.runs}
\end{table*}

\begin{figure}
\centering
\includegraphics[width=\columnwidth]{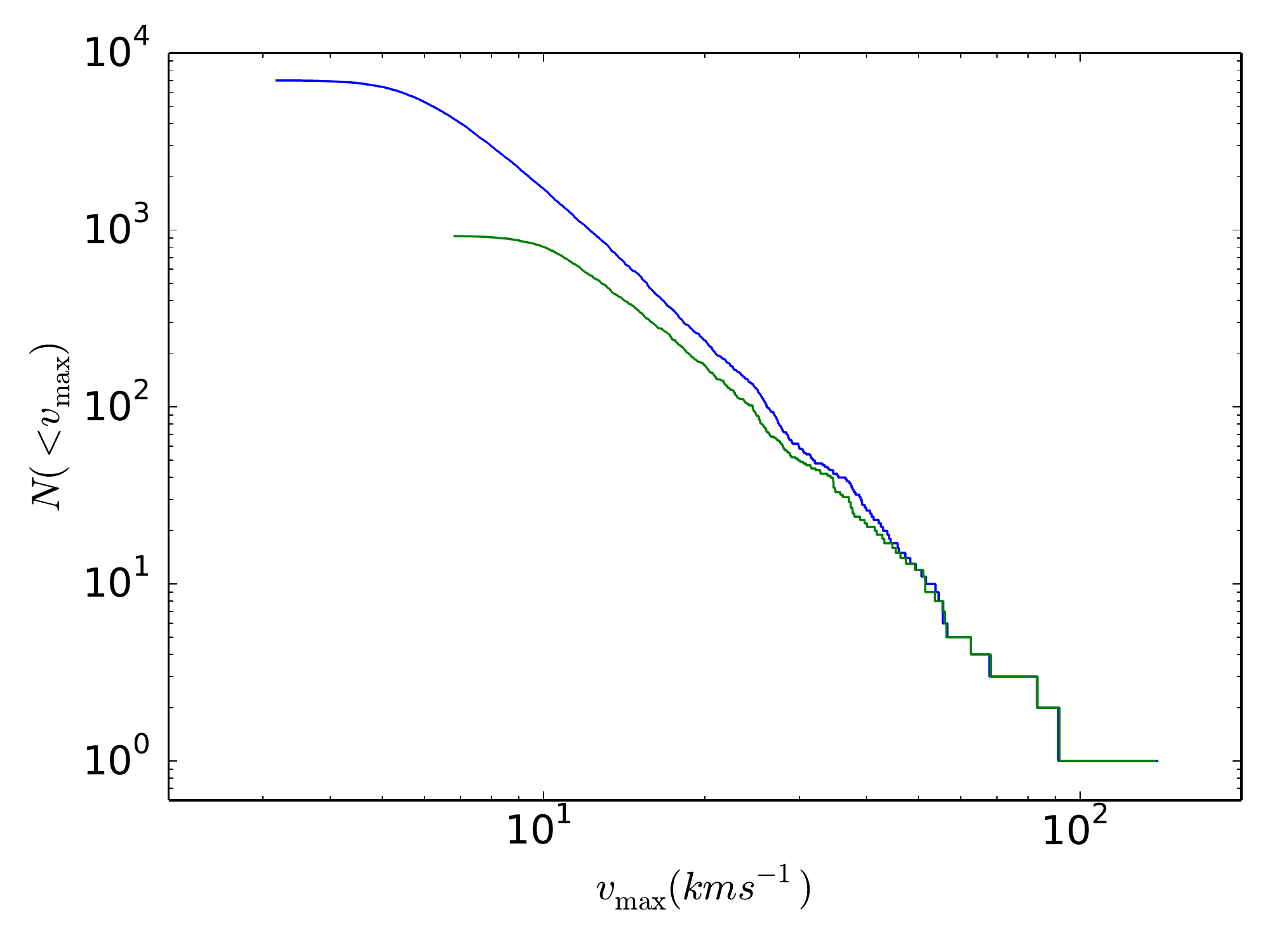}
\caption{Velocity function of the \cena\ sub-halos at $z\!=\!0$ for the 100M\_8192 (blue curve) and 100M\_4096 run (green curve).}
\label{FIG.CenANsat}
\end{figure}

Before studying the satellite population of our \cena\ analog halo in detail, we ensure our simulations are numerically convergent.~We plot the velocity function of the \cena\ sub-halos for both re-simulations in Figure~\ref{FIG.CenANsat} and find our simulation to be convergent to the resolution limit.~While the low-resolution run (100M\_4096) resolves halos down to circular velocities of $v_{\rm max}\!\approx\!10$\,km s$^{-1}$, in our higher-resolution run (100M\_8192) we resolve approximately 7000 subhalos to $v_{\rm max}\!\approx\!5$\,km s$^{-1}$.

\section{Constraints from the Dark Star Clusters}
\label{SEC.constraints}
\subsection{General Considerations}
As stated in \S~\ref{SEC.intro}, we are determining whether the DSCs can be explained by a past or present subset of dark matter subhalos around \cena. The next step in our analysis is to determine what, if any, constraints the measured baryonic properties of these systems can place on the dark matter halo characteristics.

Under the assumption that the observed stellar velocity dispersion values are unbiased measurements of the stellar velocity dispersion\footnote{For a detailed discussion on statistical and systematic uncertainties of the velocity dispersion measurements, see \cite{tay15}.}, we estimate the central mass density for each DSC that is not accounted for by the baryonic component.~In the framework of our model, we investigate whether the DSCs can be explained by the presence of past and/or present-day dark matter.~Note, in our analysis, we consider only two mass components, the observed stellar mass and a ``dark'' component which that we presume to be {\it i)} a central cusp of dark matter and/or {\it ii)} a massive central black hole. 

\subsection{Observational Constraints}
The main constraint for our model comes from the observed stellar properties of the DSCs.~We know that, at $z\!=\!0$, they have $>\!10^5 M_\odot$ of stars with high central stellar densities. GCs and ultra-faint dwarfs \citep{Brownetal:14} are among the oldest stellar systems in the universe. It has been proposed that a fraction of the ultra-faint dwarfs are fossils of the first galaxies \citep{BovillR:09, BovillR:11a, BovillR:11b}. In this work we explore whether the DSCs could be a new category of ancient stellar systems which also formed before reionization, but with a much higher stellar density than the ultra-faint dwarfs.~ We assume that, like, the Milky Way GCs, they formed the majority of their stars $>12$~Gyr ago \citep{KatzR:13}. During this epoch of the first galaxies, the majority of stars formed in dense, globular cluster like, environments \citep{Calvietal:14,Ricotti:02,KatzR:13}.

~If the local star formation efficiency within the cluster is $f_\star^{\rm local}\!\ga\!0.35$ and/or the cluster forms $N_\star\!>\!10^6$ stars, the cluster will survive the expulsion of its initial gas reservoir and remain bound for a Hubble time \citep[e.g.][]{par09, lam10, elm14}. Given ${\cal M}_{z=0}/{\cal M}_{initial} = 9.1$ \citep{PrietoG:08}, at least $\sim10$\% of the $z=0$ population of the DSCs formed in an initial, dense burst at high redshift.

Assuming these systems form stars with a stellar IMF with a characteristic mass of $1\,M_\odot$, this requires there to be at least $10^{5.5} M_\odot$ of gas in the inner $\sim\!20$\,pc of the dark matter halo.~The $\sim20$~pc scale is chosen to corresopnd to the physical scale of the DSCs at $z=0$.~Note, that this is a lower estimate since we can expect a large fraction of the stars to be stripped when the tidal radius reaches the stellar population, after the accretion into a more massive halo \citep[e.g.][]{Penarrubiaetal:08, smi13, smi15}.

While we do not assume that DSCs formed all of their stars at high redshift, we do assume, in contrast with the progenitors of the ultra faint dwarfs, that their initial burst of star formation was highly concentrated. This is in line with recent work on the formation of the first galaxies which suggests that while many of the first galaxies formed diffuse stellar systems \citep{Ricottietal:02a, Ricottietal:02b, Ricottietal:08}, a subset formed denser stellar populations \citep{Pawliketal:11,Pawliketal:13}.

\subsection{Dark Matter Halo Framework}
The need for $\sim\!10^5\!-\!10^6\,M_\odot$ of gas in the halo core gives us an estimate for the approximate dark matter mass within 20\,pc required in the DSC progenitors at early times.~Assuming the initial gas fraction of the halos follows the cosmic baryon fraction, $f_b$, we can estimate the dark matter mass in the core, ${\cal M}_{\rm DM,core}\!\approx\! {\cal M}_g/f_b$. From this required ${\cal M}_{\rm DM,core}$, we approximate the required virialization redshift, $z_{\rm vir}$, and its mass at virialization, ${\cal M}_{\rm vir}(z_{\rm vir})$.~Assuming the halos approximate an NFW profile \citep{nav96} at virialization:

\begin{equation}
\rho_{\rm NFW}(r) = {\rho_o}\left\{\frac{r}{R_s}\left(1+\frac{r}{R_s}\right)^2\right\}^{-1}
\end{equation}

\noindent where $R_s$ is the scale radius of the halo and $\rho_o$ the central density.~$\rho_{\rm NFW}$ can be written solely as a function of $z_{\rm vir}$ and ${\cal M}_{\rm vir}(z_{\rm vir})$

\begin{equation}
\rho_{\rm NFW}\left\{r,z_{\rm vir},{\cal M}_{\rm vir}(z_{\rm vir})\right\}
\end{equation}

\noindent since

\begin{equation}
3\cdot10^3 \times \Omega_M(1+z_{\rm vir})^3 = \Delta_c \left(\frac{c^3}{\ln(1+c) + \frac{c}{1+c}}\right)
\end{equation}

\noindent where c is the concentration and $\Delta_c$ is the overdensity required for halo collapse as a function of the halo's concentration, and

\begin{equation}
R_{\rm vir} = 1.5\,{\rm kpc}\left(\frac{\Omega_Mh^2}{0.147}\right)^{-1/3}\left(\frac{{\cal M}_{\rm vir}}{10^8 M_\odot}\right)^{1/3}\left(\frac{10}{1+z_{\rm vir}}\right).
\end{equation}

If we assume a spherically symmetric dark matter distribution, we obtain the dark matter core mass:
\begin{equation}
{\cal M}_{\rm DM,core} = 4\pi\rho_oR_s^3\left[\ln\left(\frac{R_s + R_{\rm core}}{R_s}\right) - \frac{R_{\rm core}}{R_s + R_{\rm core}}\right]
\label{EQ.Mcore}
\end{equation}
where $R_{\rm core}$ is the radius of the core, here assumed to be $20$~pc. 

\begin{figure}
\centering
\includegraphics[width=9cm]{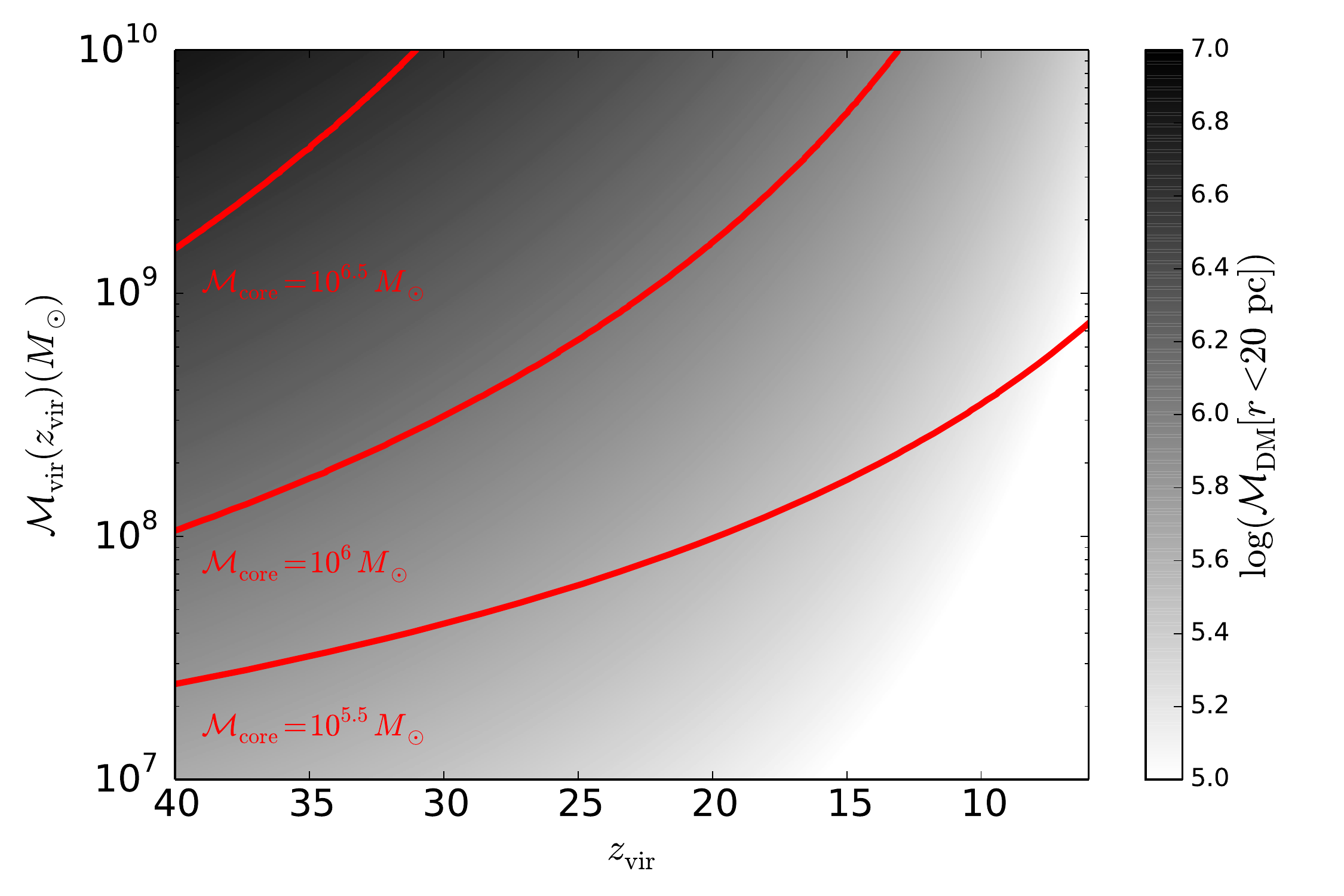}
\caption{Contour plot of the mass of dark matter within $20$\,pc for a given $z_{\rm vir}$ and ${\cal M}_{\rm vir}(z_{\rm vir})$. The range of masses on the greyscale run from $10^5 M_\odot$ to $10^7 M_\odot$ straddling our approximate value for the core dark matter mass required to form a DSC progenitor of $10^6 M_\odot$.}
\label{FIG.innermass}
\end{figure}

Using Equation~\ref{EQ.Mcore}, we plot the contours of ${\cal M}_{\rm DM,core}$ as a function of $z_{\rm vir}$ and ${\cal M}_{\rm vir}(z_{\rm vir})$ in Figure~\ref{FIG.innermass}, where we show contours of equivalent ${\cal M}_{\rm DM}(r\!<\!20\,{\rm pc})$ in greyscale.~We also plot the ${\cal M}_{\rm vir}(z_{\rm vir})$ curves for ${\cal M}_{\rm DM}(r\!<\!20\,{\rm pc})\!=\![10^{5.5} M_\odot, 10^6 M_\odot, 10^{6.5} M_\odot]$.~Note that dark matter halos with ${\cal M}_{\rm DM}(r\!<\!20 \,{\rm pc})\!\simeq\!10^6 M_\odot$ would form $\sim\!10^{5.5} M_\odot$ of stars at $f^{\rm local}_\star\!=\!0.35$, making them plausible progenitors of the DSCs. From these curves we estimate ${\cal M}_{\rm vir}(z=z_{vir})\!>\!10^8 M_\odot$, which is in agreement with the masses required for the formation of $\sim10^6 M_\odot$ proto-GC systems \citep{KatzR:13}

From the ${\cal M}_{\rm vir}(z_{\rm vir})$ required to produce the necessary ${\cal M}_{\rm DM,core}$ and given the growth of halos before infall into \cena, we assume any possible progenitor of a DSC reached a peak mass ${\cal M}_{\rm vir}(z_{\rm peak})\!>\!10^9 M_\odot$ {\it before} falling into \cena, where ${\cal M}_{\rm vir}(z_{\rm peak})$ is the maximum virial mass of a halo. 

\begin{figure}
\centering
\includegraphics[width=7.9cm]{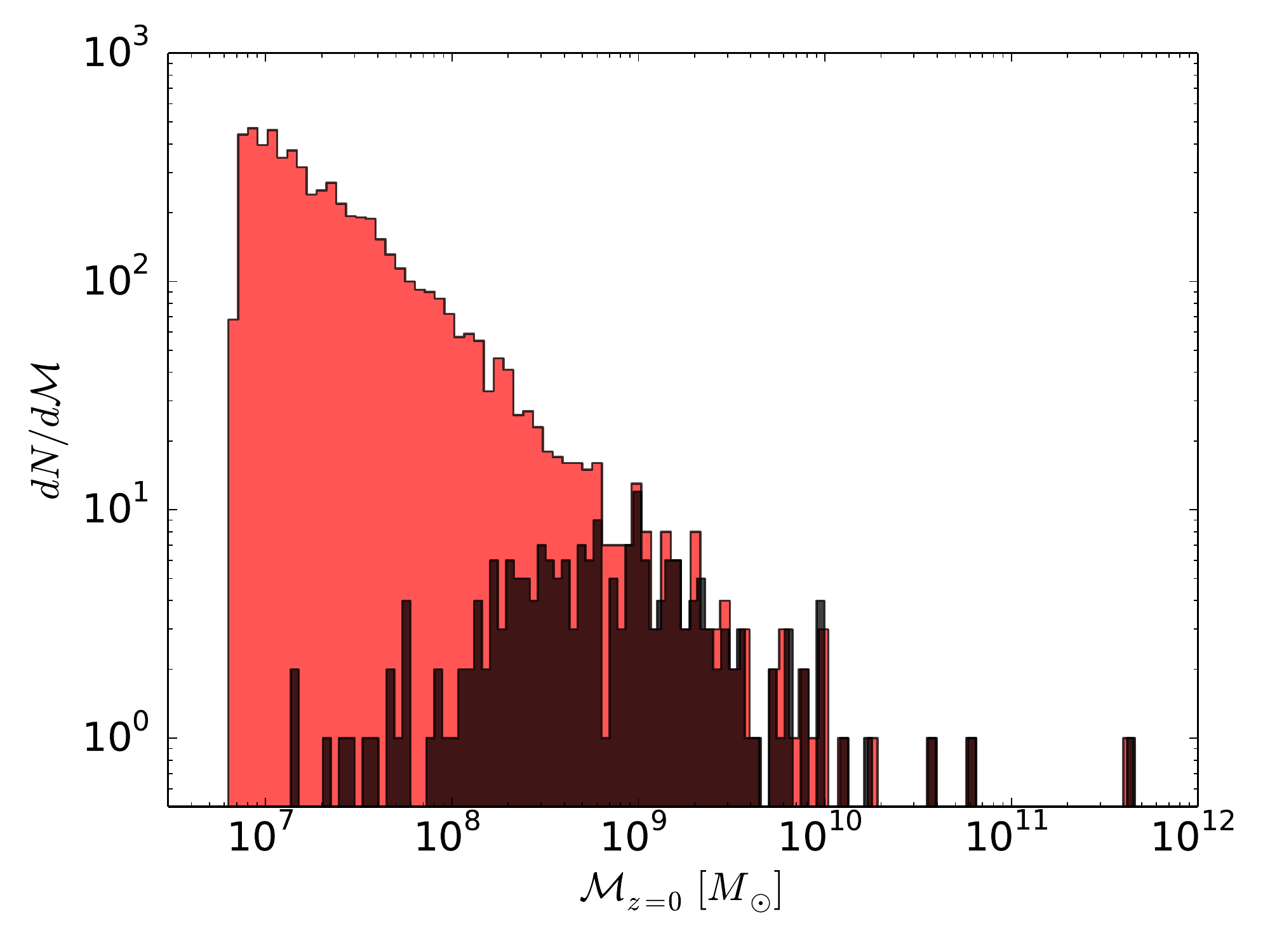}
\includegraphics[width=7.9cm]{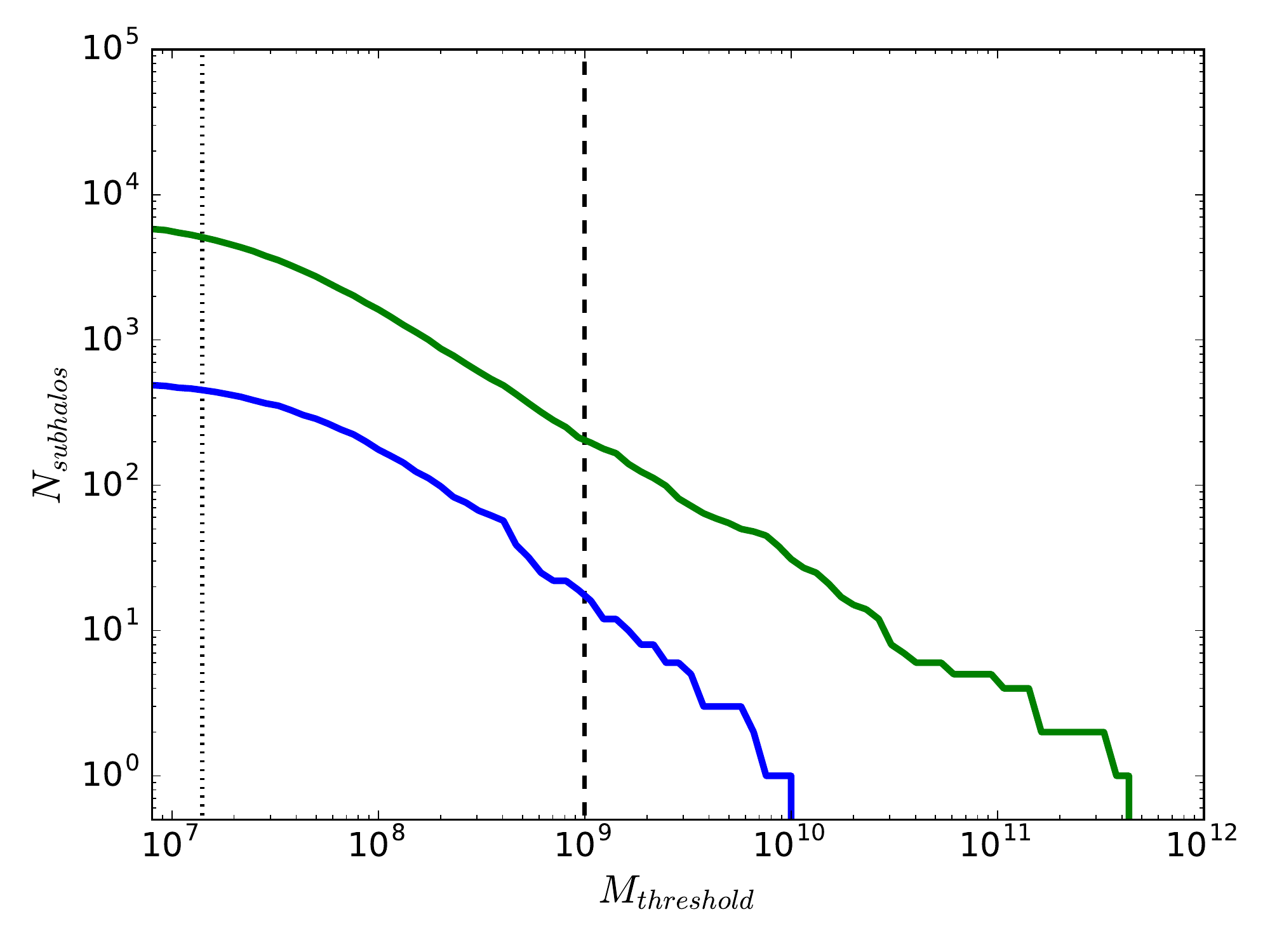}
\caption{{\it{Left}}: Mass function of all the resolved subhalos at $z\!=\!0$ (red) and the subhalos with ${\cal M}_{\rm vir}(z_{\rm peak})>10^9 M_\odot$ (black), which are considered to be progenitor halos of DSCs discovered in \cena. {\it{Right}}: Number of subhalos, $N_{subhalos}$ with a $M_{vir}(z_{peak})>M_{threshold}$ in the virial halo of our \cena\ analog (green) and $N_{subhalos}$ within $100$~kpc, projected, of the central halo (blue) for a range of $M_{threshold}$. The mass of halos resolved with $100$ particles in 100M\_8192 is shown as the dotted line and the $M_{vir}(z_{peak}) > 10^9 M_\odot$ determined from observational constraints is the dashed line.}
\label{FIG.mf}
\end{figure}

The $z=0$ mass function of all resolved \cena\ subhalos in 100M\_8192 is shown in the left panel of Figure~\ref{FIG.mf} in red, with the mass function of the subset of subhalos with ${\cal M}_{\rm vir}(z_{\rm peak})\!>\!10^9 M_\odot$ overlaid in black.~These latter halos are considered potential progenitors of the DSCs found by T15.

The number of $z=0$ subhalos with $M_{vir}(z_{peak})>M_{threshold}$ is shown in the right panel of Figure~\ref{FIG.mf}. We plot both the total number of subahlos, $N_{subhalos}$ with $M_{vir}(z_{peak})>M_{threshold}$ in the \cena\ analog`s virial halo and $N_{subhalos}$ for a given $M_{threshold}$ within $100$~kpc projected into the plane of the sky. The projected distance of $100$~kpc is chosen because all the DSCs are within $100$~kpc of \cena\ (T15). We find $\sim 20$ $z=0$ subahlos with $M_{vir}(z_{peak})>10^9 M_\odot$ within $100$~kpc projected of \cena\ and $\sim200-300$ within the entire virial halo. Please note two caveats to these numbers. First, we do not include subhalos which are not detected by Amiga at $z=0$. Such halos are assumed to be completely destroyed and we have no way of determining their positions relative to the central galaxies at $z=0$. Secondly, in the outer regions of the halo, the $z=0$ subhalos must not only account for a potential additional population of DSCs, but the known dwarf satellites of \cena\ \citep{kar07,van00,crn14,crn16} as well. 

\section{Dark Matter Halo Profiles and Central Densities}
\label{SEC.DM}
We have established that the likely DSC progenitors have ${\cal M}_{\rm vir}(z_{\rm peak})>10^9 M_\odot$.~However the DSCs have extremely high derived central densities, which naturally raises the question: do the \cena\ subhalos with ${\cal M}_{\rm vir}(z_{\rm peak})>10^9 M_\odot$ have such high central densities at $z=0$?

\subsubsection{Profile Properties at various Redshifts}
\begin{figure}
\centering
\includegraphics[width=7.9cm]{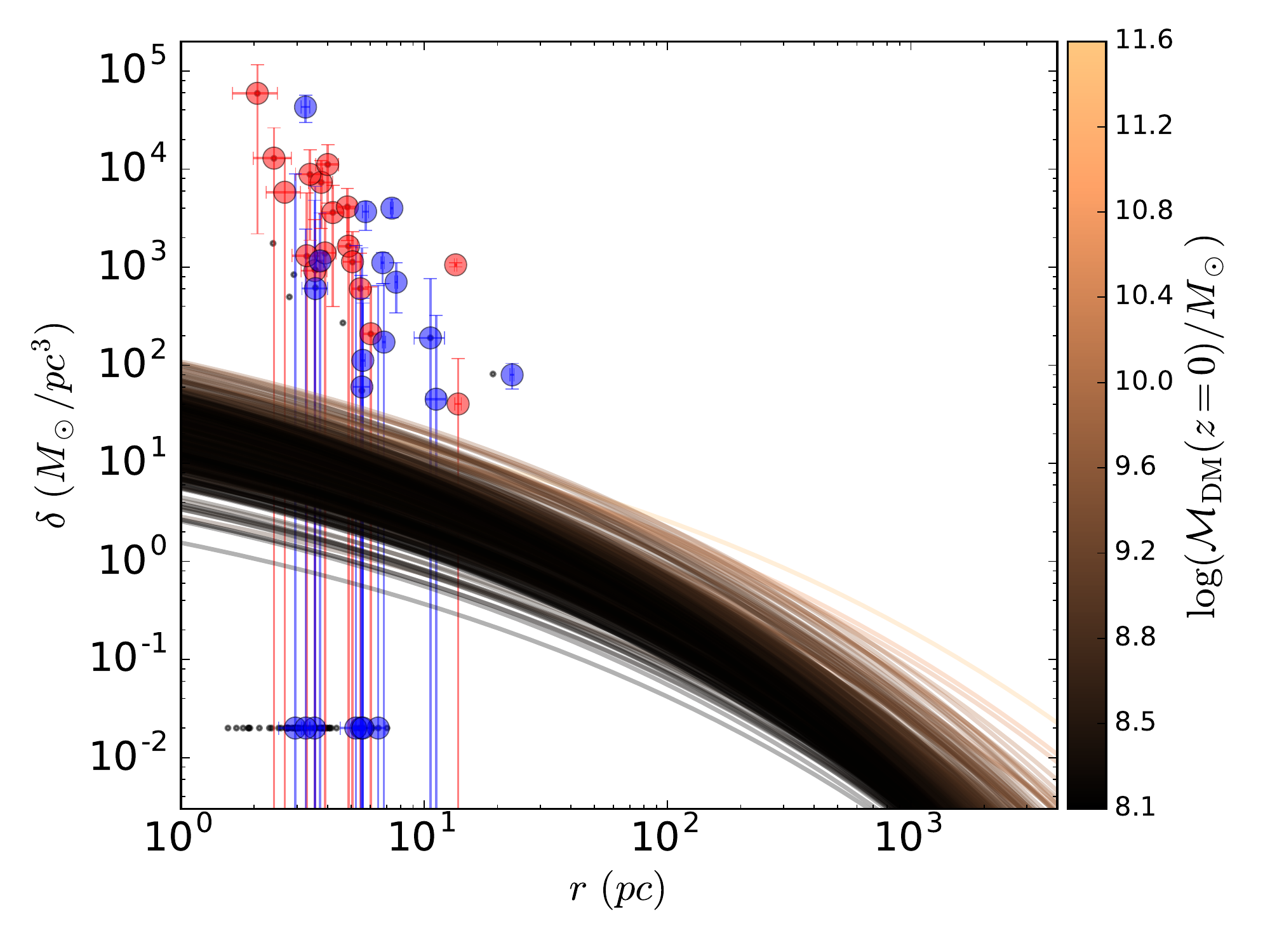}
\includegraphics[width=7.9cm]{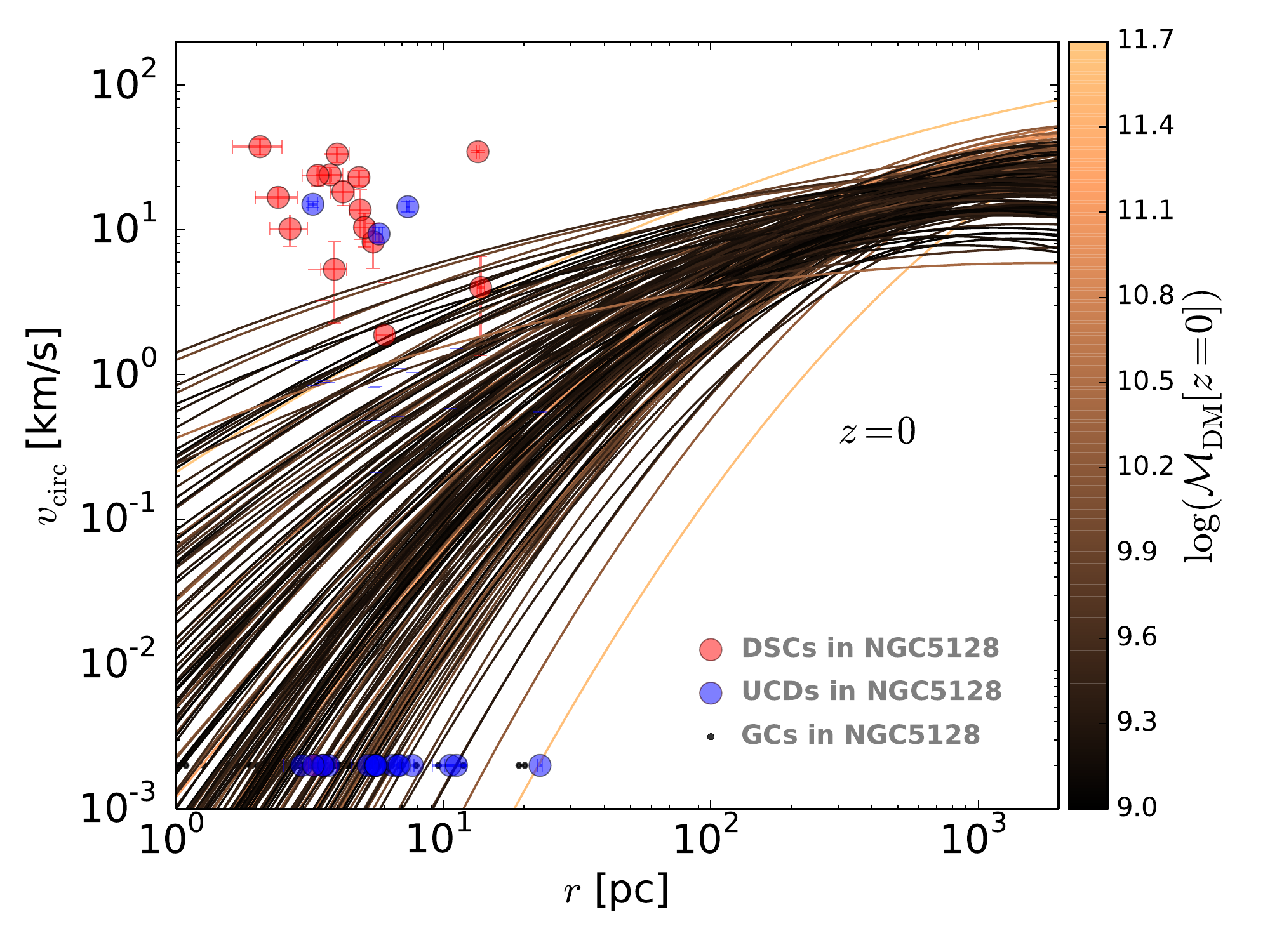}
\caption{{\it{Left}}: Einasto halo profiles for the $z\!=\!0$ subhalos of \cena\ with ${\cal M}_{\rm vir}(z_{\rm peak})>10^9 M_\odot$ with  the derived central densities for the DSCs (red symbols), UCDs (blue symbols), and GCs (black dots). The density profile colour of each halo is parametrized by the halo mass at $z\!=\!0$ (${\cal M}_{\rm DM}(z=0)$) as indicated by the colour scale on the right side of the panel. {\it{Right}}: Stoehr profiles for the subhalos of the \cena\ analog with the $\sigma_{r_{1/2}}$ versus $r_{hl}$ of the DSCs, UCDs and GCs of \cena\. The Stoehr profiles and DSCs, UCDs and GCs have the same colors as symbols as for the Einastro profiles.}
\label{FIG.z0}
\end{figure}

Our simulations do not have the resolution to directly measure the subhalo densities in the inner $20$~pc. We therefore fit a halo profile to the subhalos at various redshifts and extrapolate an estimate of the density from the simulation convergence radius at $\sim600$ kpc/h to 20 pc.~We present a detailed discussion of the convergence of our profiles in Appendix~A.~Since NFW profiles are not accurate fits for subhalos \citep{Stoehr:06}, we use the Einasto profiles \citep{Einasto:65,Navarroetal:04} to fit subhalo profiles at $z=0$. The Einasto profile approximates the radial dark matter density profile with the following analytical form:
\begin{equation}
\rho(r) = \rho_{-2} e^{ \frac{-2}{\alpha} \left[\left(\frac{r}{r_{-2}}\right)^{\alpha}-1.0\right]},
\end{equation}
where $r_{-2}$ is the radius at which the slope is -2, $\rho_{-2}$ is the density at that point and $\alpha$ is between 0.16 and 0.20 \citep{Springeletal:08}. For $\alpha=0.18$ this becomes,
\begin{equation}
\rho(r) = 99.5 M_\odot pc^{-3} \frac{v_{max}}{r_{max}}^2 e^{ \frac{-2}{\alpha} \left[\left(\frac{2.189 r}{r_{max}}\right)^{\alpha}-1.0\right]} ,
\end{equation}
where $r_{max}$ and $v_{max}$ are halo parameters derived by Amiga. In Figure~\ref{FIG.z0}, we plot the Einasto density profiles fit to the subhalos with ${\cal M}_{\rm vir}(z_{\rm peak})\!>\!10^9 M_\odot$ and compare them to the derived densities for the various star cluster populations in \cena\ (see T15).~The Einasto profiles are shaded according to their $z=0$ dark matter mass.~We find that the even the densest subhalos in our simulation at $z\!=\!0$ have central densities more than an order of magnitude too low to account for the average DSC.

We also show the Stoehr profiles \citep{Stoehretal:02} to fit subhalo profiles at $z=0$. The Stoehr profile approximates the circular velocity using:
\begin{equation}
\log\left(\frac{v_c}{v_{\rm max}}\right) = -a \left(\log\left\{\frac{r}{r_{\rm max}}\right\}\right)^2,
\end{equation}
where $a$ is a parameter fit to the resolved portions of the circular velocity curve calculated by {\sc Amiga}. In Figure~\ref{FIG.z0}, we plot the circular velocity curves of the Stoehr profile fits for the subhalos with ${\cal M}_{\rm vir}(z_{\rm peak})\!>\!10^9 M_\odot$ and compare to the line-of-sight radial velocities for the various star cluster populations in \cena\ (see T15).~The Stoehr profiles are shaded according to their $z=0$ dark matter mass.~Although some Stoehr profiles with the highest central densities match some of the lowest-density DSCs, we find that, as with the Einasto profiles, the subhalos in our simulation at $z\!=\!0$ have central densities more than an order of magnitude too low to account for the average DSC. As such, the inability of dark matter alone to account for the high derived densities of the DSC is independent of our choice of subhalo profile.

\begin{figure*}
\centering
\includegraphics[width=7.9cm]{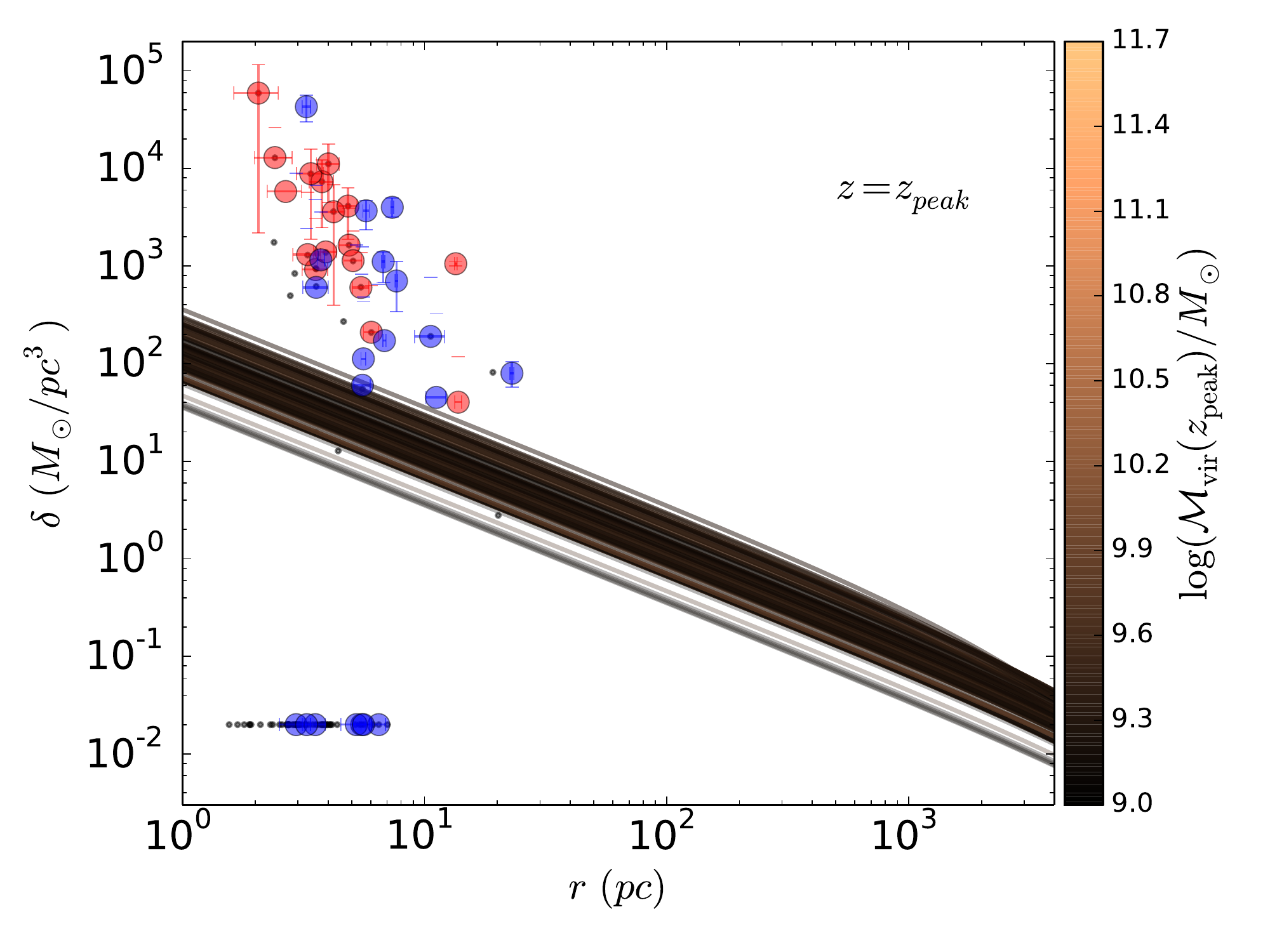}
\includegraphics[width=7.9cm]{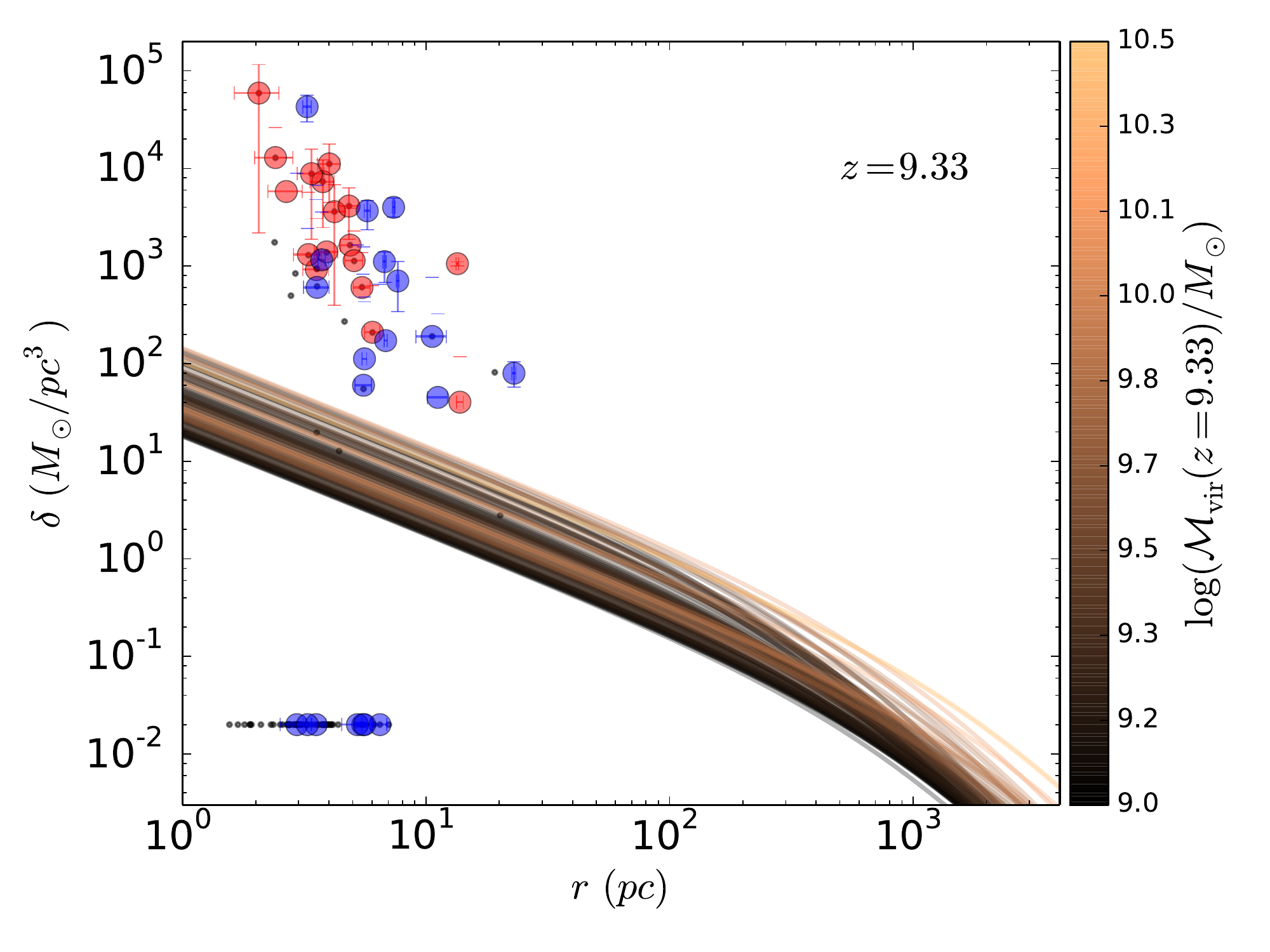}
\caption{({\it Left panel}): NFW profile fits based on the {\sc Amiga} parameters derived at $z_{\rm peak}$ for each $z=0$ \cena\ subhalos with ${\cal M}_{\rm vir}(z_{\rm peak})>10^9 M_\odot$ with the derived central density of the DSCs at $z=0$.~({\it Right panel}): NFW profiles of the \cena\ progenitors at $z=9.3$. The NFW profiles are color coded with ${\cal M}_{\rm vir}(z)$ at the corresponding redshifts.~The central densities derived for the \cena\ star cluster population are as in Figure~\ref{FIG.z0}.}
\label{FIG.zpeak}
\end{figure*}

However, we are not only interested in the dark matter density at $z\!=\!0$, but also in the central density of halos at $z_{\rm peak}$, when these were still actively forming stars, before their evolution became dominated by \cena.~In the left panel of Figure~\ref{FIG.zpeak}, we plot the NFW profiles of the halos at $z_{\rm peak}$. In both panels of Figure~\ref{FIG.zpeak}, we use the same symbol colors as for the data points in Figure~\ref{FIG.z0}, however, the NFW profiles in both panels are shaded by ${\cal M}_{\rm vir}(z_{\rm peak})$ rather than ${\cal M}_{\rm DM}(z=0)$.

Before stating categorically that dark matter alone cannot account for the densities of a significant fraction of the DSCs, we check one final set of high-redshift subhalos.~The progenitors of densest star clusters in \cena\ likely formed at high redshifts. As redshift increases so does the fraction of progenitors which do not survive as $z\!=\!0$ subhalos.~However, the right panel of Figure~\ref{FIG.zpeak} shows that, even at high redshift (e.g. $z\!=\!9.3$), there are no \cena\ progenitors or $z\!=\!0$ subhalos which have dark matter central densities high enough to explain the average observed central densities of most observed star clusters from the T15 sample.

\subsubsection{Excess Central Densities}
The critical difference between the DSCs, UCDs and GCs is clear upon further study of Figures~\ref{FIG.z0}~\&~\ref{FIG.zpeak}. For all these populations, we assume a stellar \mlr\ ratio (\smlr) to estimate the stellar mass of each system. The extensively studied Milky Way GC system has a typical \smlr$\simeq\!2.2$.~Comprehensive studies of the \smlr\ values in the \cena\ system currently do not exist.~We therefore estimate the typical \smlr\ for the \cena\ GC system by measuring the peak of the ${\cal M}_{\rm dyn}/{\cal L}$ distribution for the GCs (black points), and approximate \smlr$\sim\!4.0$. In this idealized scenario, we assume the mass of the \cena\ GCs is entirely accounted for by their stellar populations. 

In the following, each density excess measure is derived using
\begin{equation}
\rho_{\rm excess} = \rho_{\rm dyn} - \rho_{\rm baryon} = \frac{{\cal M}_{\rm dyn} -  {\cal L}({\cal M}/{\cal L}_\star)}{ \frac{4}{3} \pi r_e^3}
\end{equation}
where ${\cal M}_{\rm dyn}$ is computed from the observed $\sigma_\star$, ${\cal L}$ is the derived V-band luminosity assuming a distance modulus, $m-M=27.88$ \citep{har10}, $r_e$ is the effective radius and \smlr\ is the stellar mass-to-light ratio in the V-band. For visualization purposes, if an object has $\rho_{\rm excess} < 0$ we assign an arbitrary $\rho_{\rm excess} = 2\times10^{-2}$. Such objects are not unphysical but have \smlr$<\!4.0$. With a reasonable \smlr, the dynamical mass of a significant fraction of the GCs and UCDs can be accounted for with their stars alone. This is not the case for the DSCs. While there is evidence that stellar populations in elliptical galaxies and UCDs have bottom heavy IMFs, for both UCDs and ellipticals the bottom heavy IMFs increase the stellar $M/L$ by no more than a factor of five \citep{Cappellarietal:12,MieskeK:08,Dabringhausenetal:08}. This is not enough to account for the measured dynamical $M/L$s of the DSCs ($10-100$).

We find that dark matter alone cannot account for the high observed central densities ($\rho_{\rm excess}\!\ga\!\rho_{\rm DM}$, see Figures~\ref{FIG.z0}~\&~\ref{FIG.zpeak}).~Assuming the observations are correct, the only physical possibility to explain $10^5\!-\!10^7\,M_\odot$ of mass within $\sim\!10$\,pc radius is a central black hole in the SMBH range.~T15 assumed that the DSCs fall on the ${\cal M}_\bullet$-$\sigma$ relation.~However, the validity of this assumption for tidally stripped systems is questionable. In the subsequent analysis, we will not make this assumption, but rather derive the required ${\cal M}_\bullet$ to account for the missing central mass density of our subhalos ($\rho_{\rm excess}$) that cannot be accounted for by the DM density profiles and the stellar mass distribution.~We estimate the central black-hole mass of the clusters with either ${\cal M}_\bullet = {\cal M}_{\rm dyn} - {\cal M}_\star - {\cal M}_{\rm DM}$ or ${\cal M}_\bullet = {\cal M}_{\rm dyn} - {\cal M}_\star$. In the former, we assume a cusp of the original dark matter halo has survived and, in the latter, we assume a purely baryonic stellar system at $z\!=\!0$. The difference is negligible as ${\cal M}_{\rm DM}(r<\!10\, {\rm pc}) \ll {\cal M}_{\rm dyn}$ for even the densest halos. 


We find the central black hole must have a mass in the range ${\cal M}_\bullet\!\approx\!10^5\!-\!10^7 M_\odot$, well above the estimates for IMBH candidates in GCs \cite[e.g.][]{lue13}, however this is not beyond the range seen in dwarf galaxies \citep{ReinesV:15}.~Therefore, the embedding of DSC progenitors in DM halos remains a critical component of their formation and evolution.

\section{The Central Black Hole}
\label{SEC.seed}

\begin{figure}
\centering
\includegraphics[width=8.9cm]{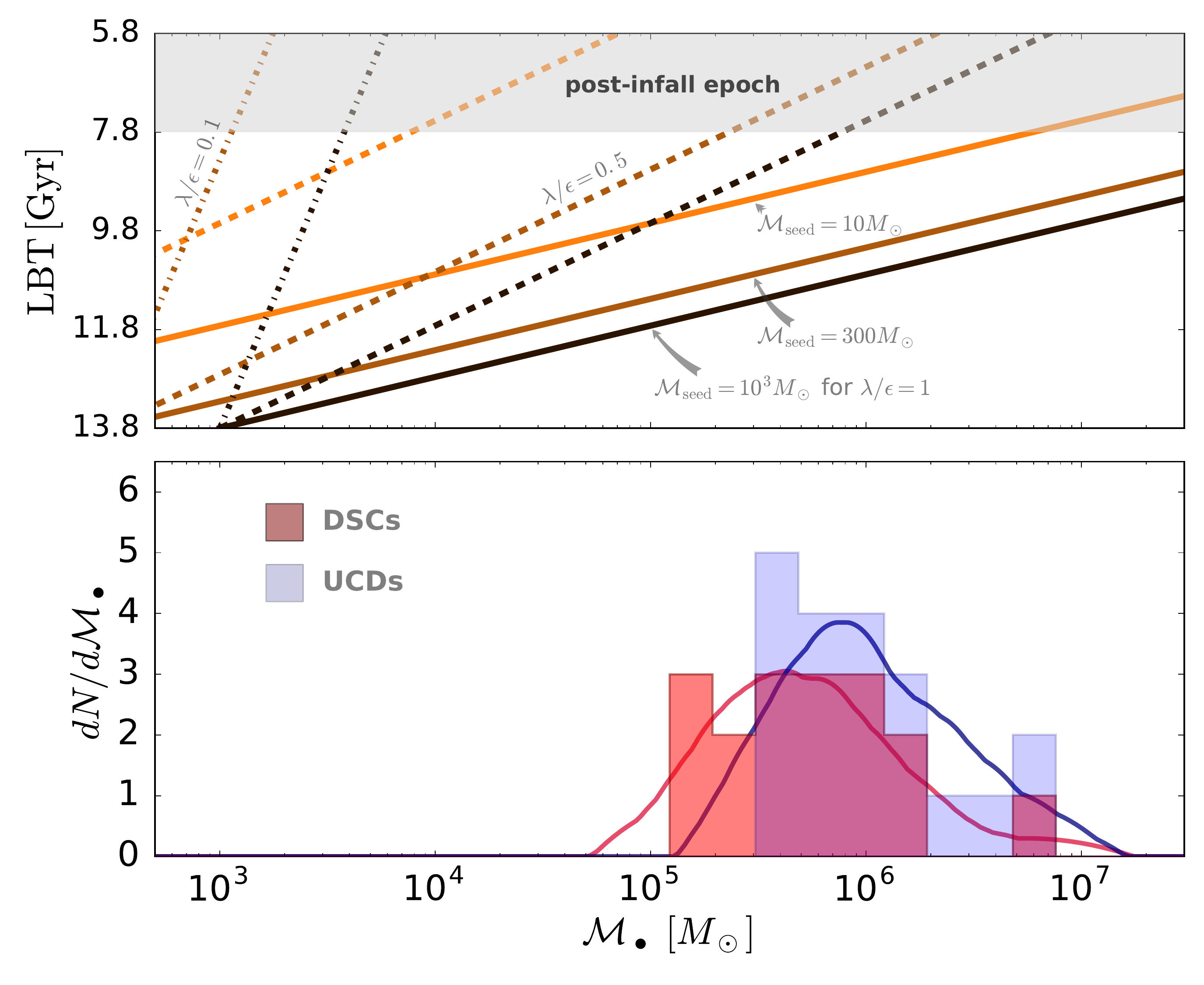}
\caption{{\it(Top panel)}:~The growth of three different seed black hole masses for a variety of $\lambda/\epsilon$. We choose ${\cal M}_{BH} = 10 M_\odot$ (light orange) to represent a CCSN remnant, ${\cal M}_{BH} = 300 M_\odot$ (dark orange) and ${\cal M}_{BH} = 10^3 M_\odot$ (black) to approximate the remnant of a runaway stellar collision. We explore three different values of $\lambda/\epsilon = [1, 0.5, 0.1]$ as the solid, dashed, and dotted lines, respectively. For thin disk accretion this corresponds to 10\%, 5\% and 1\% of Eddington accretion.~{\it (Bottom panel)}: Distributions of the central black hole masses necessary to explain the excess dynamical mass that cannot be explained by dark matter densities, derived for the high-\mlr\ stellar systems in \cena, i.e.~UCDs and DSCs (see T15).~The curves indicate Epanechnikov kernel non-parametric probability density estimates.}
\label{FIG.mbh}
\end{figure}

We established in Section~\ref{SEC.DM} that dark matter alone cannot reproduce the observed densities of the DSCs, and only a black hole in the $10^5\!-\!10^7 M_\odot$ mass range can account for the observed dynamics.~We illustrate in Figure~\ref{FIG.mbh} the distributions of the derived black hole masses for UCDs and DSCs.~Therefore, a critical component of our model is the formation and growth of a central black hole which reaches at least ${\cal M}_\bullet\!>\!10^5 M_\odot$ before the infall of the host halo into \cena. 

In the following, we assume for simplicity that the central black hole seed grows only via gas accretion at a constant accretion rate over cosmic time.~We define an accretion ratio, $\lambda/\epsilon$, which is the quotient between the Eddington ratio ($\lambda$) over the mass-energy conversion efficiency ($\epsilon$). The Eddington ratio, $\lambda\!=\!\dot{{\cal M}}_{\rm acc}/\dot{{\cal M}}_{\rm Edd}$ is the fraction of the Eddington accretion rate sustained by the central black hole.~To get a physical accretion rate in $M_\odot/$yr from $\lambda$ we need to assume an efficiency for the conversion of mass to energy, $\epsilon$, as the matter falls onto the central black hole. Since the Eddington limit is set by the luminosity, lower $\epsilon$ will allow more mass to be accreted onto the black hole for the same $\lambda$.

While the required $z\!=\!0$ black hole masses are large for GCs (see Figure~\ref{FIG.mbh}), they are on the low end of the SMBH range. We will initially look at Bondi accretion onto these systems.~The Bondi accretion rate depends on the black hole mass and the density of the surrounding gas.~For an IMBH, $\lambda_{\rm Bondi}=\!\dot{{\cal M}}_{\rm Bondi}/\dot{{\cal M}}_{\rm Edd}\!\approx\!10^{-4}$ \citep{Kongetal:10} and radiative feedback from an IMBH gives a true accretion rate of $1$\% of Bondi \citep{ParkR:11} giving a $\lambda\!\simeq\!10^{-6}$. We assume $\epsilon\!=\!0.1$, which is reasonable for thin disk accretion if we ignore black hole spin \citep{Shapiro:73,ParkO:01}.~This gives us an accretion ratio of approximately $\lambda/\epsilon\simeq10^{-5}$.

However, with Bondi accretion it is not possible to grow the central black hole to the necessary mass before infall into \cena\ unless the seed black hole is above $10^5 M_\odot$. We therefore assume the black holes are accreting well above the Bondi rate, consistent with coevolution of a BH and bulge \citep{Parketal:15}. 

Figure~\ref{FIG.mbh} illustrates the mass growth of three different seed black holes with masses $1000\,M_\odot$, $300 M_\odot$ and $10\,M_\odot$, with the assumption that the seed formed concurrently with the first generation of Pop-II stars.The $1000 M_\odot$ black hole seed approximates the remnant of a runaway stellar collision (black curves).~The $300\,M_\odot$ seed (dark orange curves) is the lower bound of the range of masses required for direct collapse \citep{heg03}.~The final set of light orange curves is for a $10\,M_\odot$ black hole seed, which approximates the growth via accretion onto a core-collapse SN remnant. Further study is needed to determine if our toy model for black hole accretion holds up to a more physically rigorous treatment.

To estimate the required accretion ratio to reach the $10^5\!-\!10^7 M_\odot$ black-hole mass range derived from observations (see histograms in Figure~\ref{FIG.mbh}), we use two criteria. First, the gas supply to the central black hole will be extinguished by ram pressure stripping after the DSC progenitor falls into \cena. Second, we know that by $z\!=\!0$ the DSC has lost the vast majority of its dark matter and stars, suggesting the near complete destruction of the DSC progenitor halo. Therefore, the seed black hole must reach the observed mass range (i.e.~$>\!10^5\, M_\odot$) by $z\!\approx\!1$ in order to allow its host halo enough time to be destroyed via interactions with \cena\ during the post-infall epoch. ~It is clear from Figure~\ref{FIG.mbh} that, as long as $\lambda/\epsilon\!\ga\!1$, it is quite possible to grow the measured black hole masses from a $300\,M_\odot$ seed and even from a $10\,M_\odot$ seed by $z\!\approx\!1$, i.e.~before the post-infall epoch begins and the environment of the black hole is evacuated by ram-pressure stripping and other interactions in the host halo.

\subsection{Observational Tests for the Model}
\label{SEC.tests}

We propose that the DSCs, stripped of their dark matter and outer stars, are remnant stellar cusps surrounding central IMBH/SMBHs. Our model for the formation of the DSCs is roughly as follows.

\begin{enumerate}[leftmargin=*]

\item During the epoch of the first galaxies a dense, massive ($M>10^5 M_\odot$) central cluster forms in a dark matter halo.\\

\item Rapid segregation of the most massive stars to the core results in a runaway stellar collision \citep{PortegiesZwartetal:04, vanderMarel:04}. This collision produces a seed black hole of several hundred to several thousand solar masses \citep{Katzetal:15}.\\
	
\item Over the next several billion years the halo continues to accrete gas, feeding additional star formation and the central black hole. During this era, feedback from bursty star formation precludes the formation of a significant bulge \citep{Guedesetal:11,Governatoetal:10}.\\

\item Between $5-10$~Gyr ago, the dwarf falls into \cena\ . This allows enough time for its gas, dark matter and outer stars are completely stripped \citep{smi13, smi15}, therefore the DSCs would not be hosted in the surviving subhalos at $z=0$. At $z=0$, the only remaining sign of the lost dwarf is the remnant cusp of stars surrounding its central SMBH. Excepting its dynamics and metallicity, the surviving stellar cusp resembles a GC.

\end{enumerate}

Since these systems resemble GCs photometrically and structurally, it is only possible to disentangle their origin as lost dwarf galaxies through the dynamic signatures of their central black holes and the following observational criteria.

{\bf{Chemical Composition}}: If the DSCs formed as dwarf galaxies their residual stellar populations should accordingly have dwarf galaxy-like chemical abundances; namely, a significantly elevated metallicity ${\rm [Fe/H]}\!>\!-2$ dex, a measurable metallicity spread ($\sigma_{[Fe/H]}>0$), and a higher [$\alpha/$Fe] ratio than expected for their present stellar mass on the [$\alpha/$Fe]-luminosity relation of stellar systems \citep[e.g.][]{tho05, smi09}. In addition, they should be above expectations from the $\rm L-[Fe/H]$ for ${\cal M}_{\star} = 10^5 - 10^6 M_\odot$ systems. From the selective core-collapse SN enrichment we would also expect enhanced light-element (CNO) to iron ratios, as well as a significant variance in iron-peak elements.

{\bf{X-ray flux}}: While the central black hole is likely now quiescent, low-level accretion ($\sim\!10^{-8}$ Eddington) is possible via AGB winds and planetary nebulae. At \cena's distance, even these low accretion rates would be detectable by Chandra or XMM for the expected central black hole masses.


\section{Summary and Conclusions}
\label{SEC.conclusions}

We have used a new set of high resolution simulations of an isolated $10^{13} M_\odot$ \cena\ analog to investigate possible formation models for the DSCs found in T15. We assume the DSCs formed within dark matter halos before determining which \cena\ subhalos at $z=0$ have the highest probability of being DSC progenitors.

We find that DSCs likely formed in dark matter halos which virialized at high redshift with ${\cal M}_{vir}(z_{vir})>10^8 M_\odot$. Since dark matter halos grow in mass in isolation, we assume that any DSC progenitor had ${\cal M}_{\rm vir}(z_{\rm peak}) > 10^9 M_\odot$. However, no dark matter profile associated with the \cena\ system, whether for the subhalos at $z=0$, $z_{peak}$, or high redshift (eg.  $z=9.3$), has a high enough central density to account for the dynamical mass derived for the DSCs. Even the densest DM halos fall one to two orders of magnitude short of explaining the DSCs as a population. Therefore, we find that dark matter alone cannot account for the currently measured dynamical masses in the DSC cores.


The only class of objects that can account for $>10^5 M_\odot$ of mass within 10 pc of the cluster center are massive central black holes. We find that the dynamical masses of the DSCs can be accounted for by central black holes between $10^5-10^7 M_\odot$. While evidence for IMBHs has been found in several globular clusters, the mass range required to explain the DSCs is several orders of magnitude higher. However, such supermassive black holes would need to form in a galaxy. Therefore, while dark matter cannot explain the $z=0$ dynamics of the DSCs, the derived central black hole masses strongly support their formation and early evolution within dark matter halos. 

Future work is required to constrain the number and distribution of these objects around \cena\ and to determine if there are likely to be any lost dwarfs masquerading as globular clusters in our own Milky Way.

\section*{Acknowledgments}
We wish to thank Nelson Padilla, Chris Reynolds and John Wise for useful discussions and Derek Richardson for help with the University of Maryland HPCC.~MSB acknowledges the support of FONDECYT Project Grant (No.~3130549).~MAT acknowledges the financial support provided by an excellence grant from the ÒVicerrector'a de Investigaci—nÓ and the Institute of Astrophysics Graduate School Fund at Pontificia Universidad Cat—lica de Chile and the European Southern Observatory Graduate Student Fellowship program. This research was supported by FONDECYT Regular Grant (No.~1161817) and BASAL Center for Astrophysics and Associated Technologies (PFB-06). The authors acknowledge the University of Maryland supercomputing resources (http://www.it.umd.edu/hpcc) made available for conducting the research reported in this paper.

\appendix

\section{Convergence Tests}
\label{SEC.appen}

With half light radii of less than $20$ pc, comparing the derived DSC densities to the dark matter halo profiles in our simulations requires us to extrapolate two orders of magnitude from the convergence radius of our simulations ($3\epsilon \sim 600$pc physical). We chose not to run simulations with a convergence radius of a few parsecs because of the computational expense and the increasing importance of baryonic effects of the dark matter distribution on sub-kiloparsec scales \citep{ReedG:05,Mashchenkoetal:06,Mashchenkoetal:08,Governatoetal:10,Governatoetal:12,Teyssieretal:13,DiCintioetal:14,PontzenG:12,Madauetal:14}. 

It is standard practice to use derived dark matter density profiles to study the densities at or the softening length of the simulation \citep{Hanetal:16,RodriguezPueblaetal:16,Gaoetal:12}. We use information derived from {\sc Amiga} to fit a dark matter density profile and extrapolate that profile into the inner $20$ pc of the halo and determine the densities of the inner dark matter profiles of our DSC candidate progenitors at various redshifts.

To test the convergence of our halo density profiles and therefore our estimates of the dark matter density within $20$ pc, we compare the density profile fits for our DSC progenitors in the 100M\_8192 and 100M\_4096 runs. The properties of these runs are summarized in Table 1 in the main text.

We remind the reader that the DSC candidates are chosen to be resolved by Amiga at $z=0$ with $M_{vir}(z_{peak})>10^9 M_\odot$. We only investigate those $z=0$ subhalos which are resolved with $N>100$ particles, though {\sc Amiga} is robust to $N>50$ \citep{KnollmanK:09}. Throughout the analysis in this appendix the curves and points for the 100M\_4096 will be shown in green and the curves and points for the 100M\_8192 will be shown in black.

\begin{figure*}
\centering
\includegraphics[width=5.8cm]{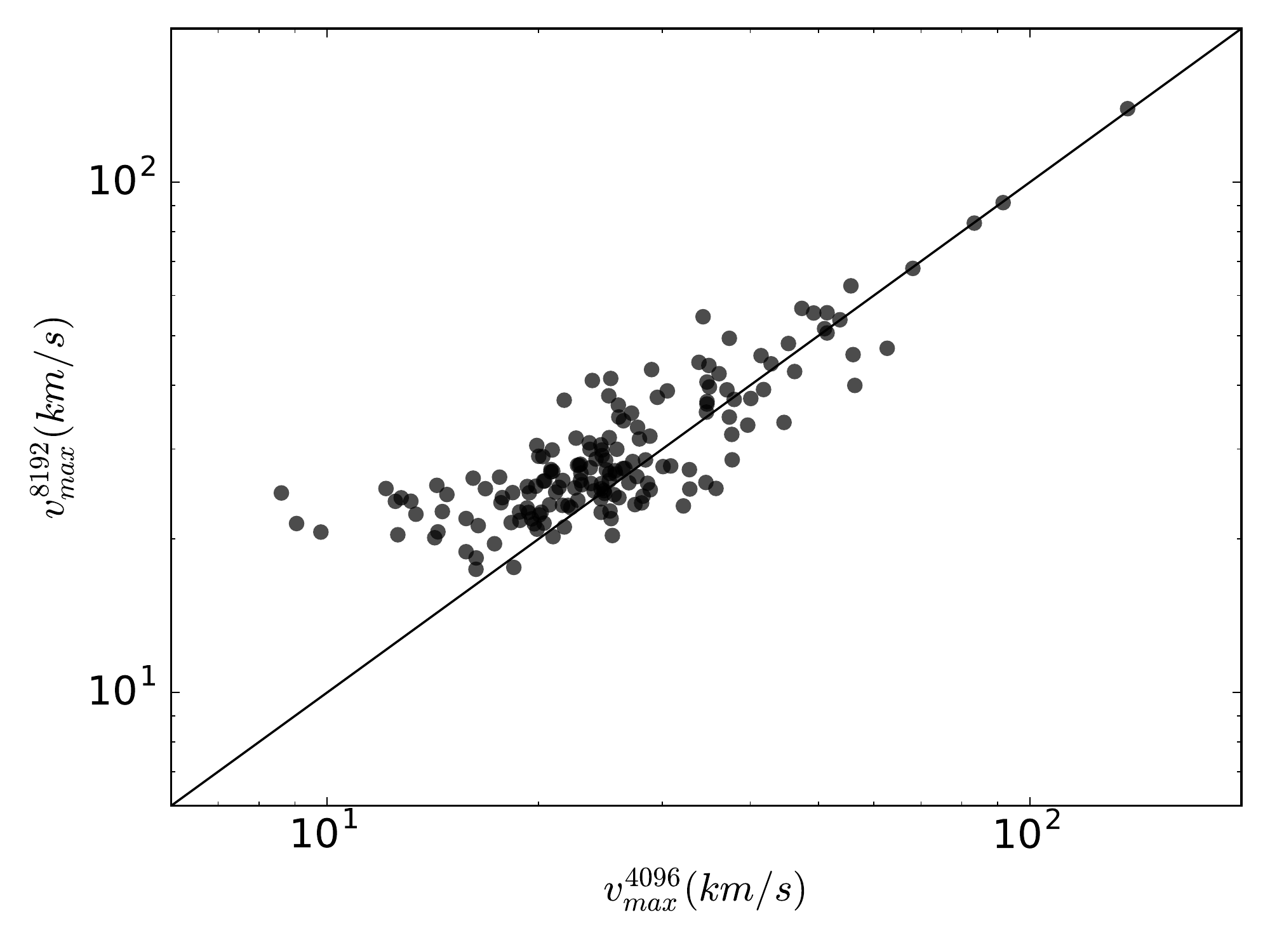}
\includegraphics[width=5.8cm]{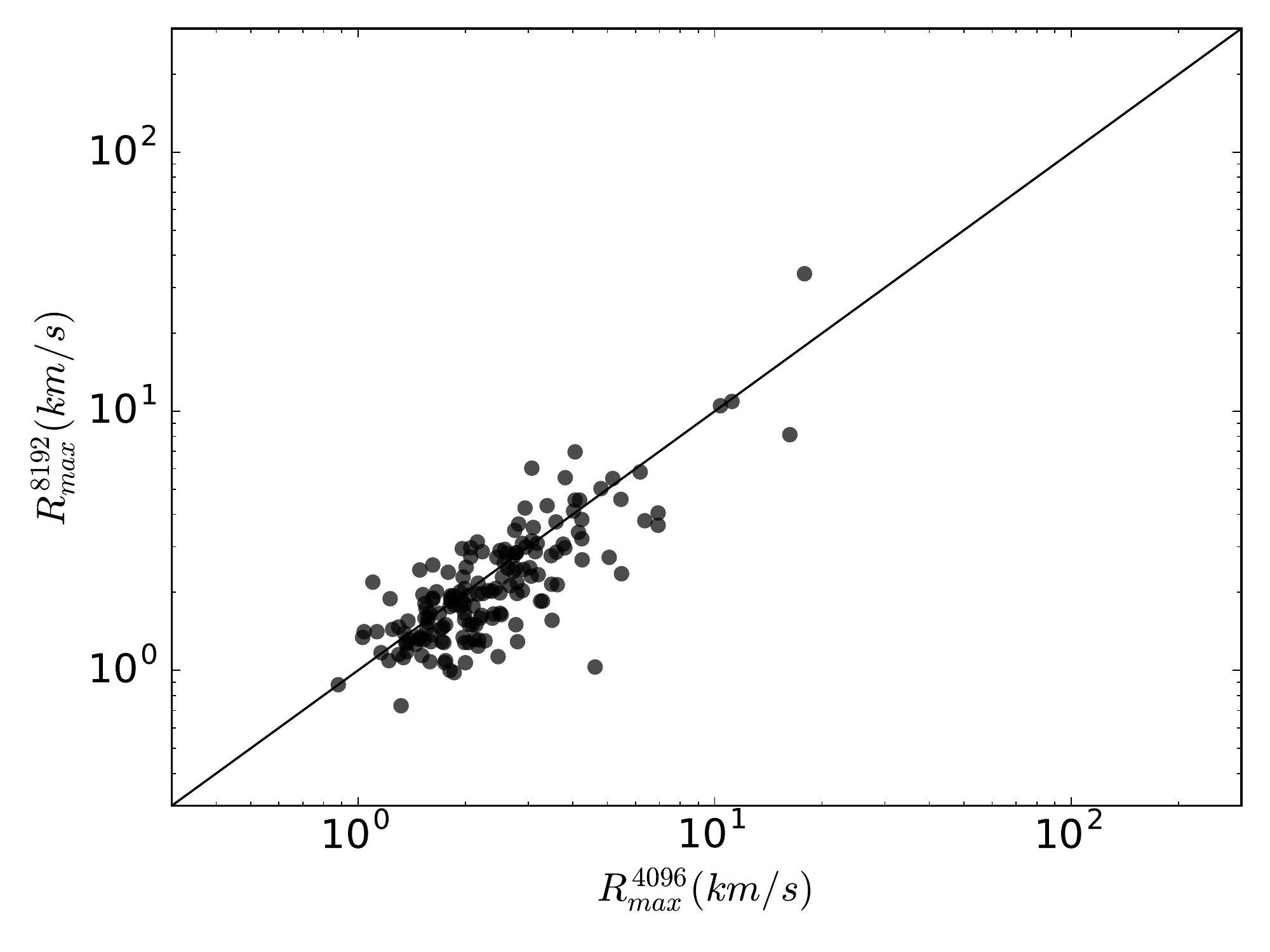}
\includegraphics[width=5.8cm]{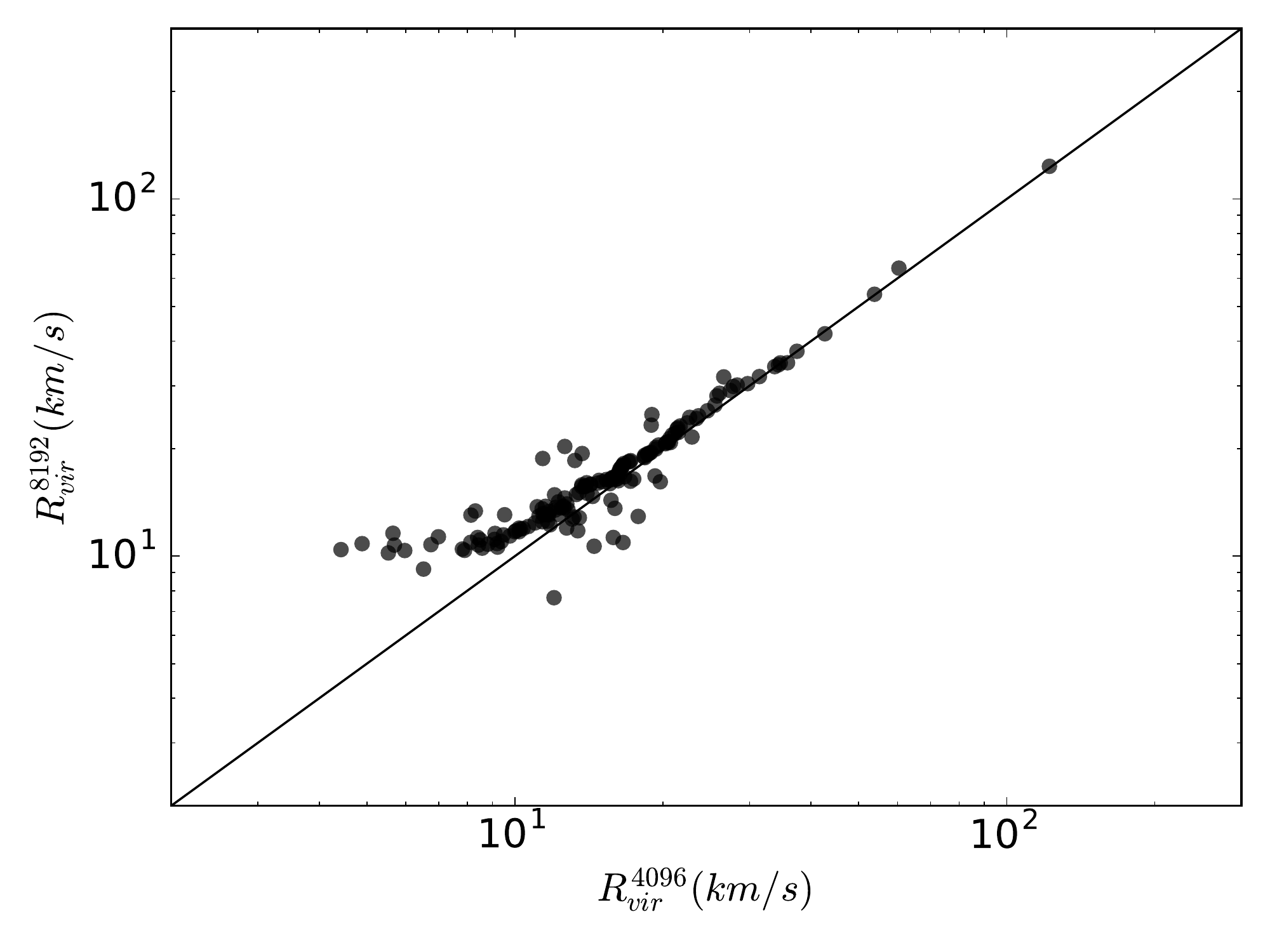}
\caption{For the DSC candidate progenitor subhalos resolved in 100M\_4096 and their counterparts in 100M\_8192 we plot $v_{max}$ (left panel), $r_{max}$ (center panel), and $r_{vir}$ (right panel). In each panel, the thin black line shows equivalence.}
\label{FIG.convergence}
\end{figure*}

\begin{figure*}
\centering
\includegraphics[width=5.8cm]{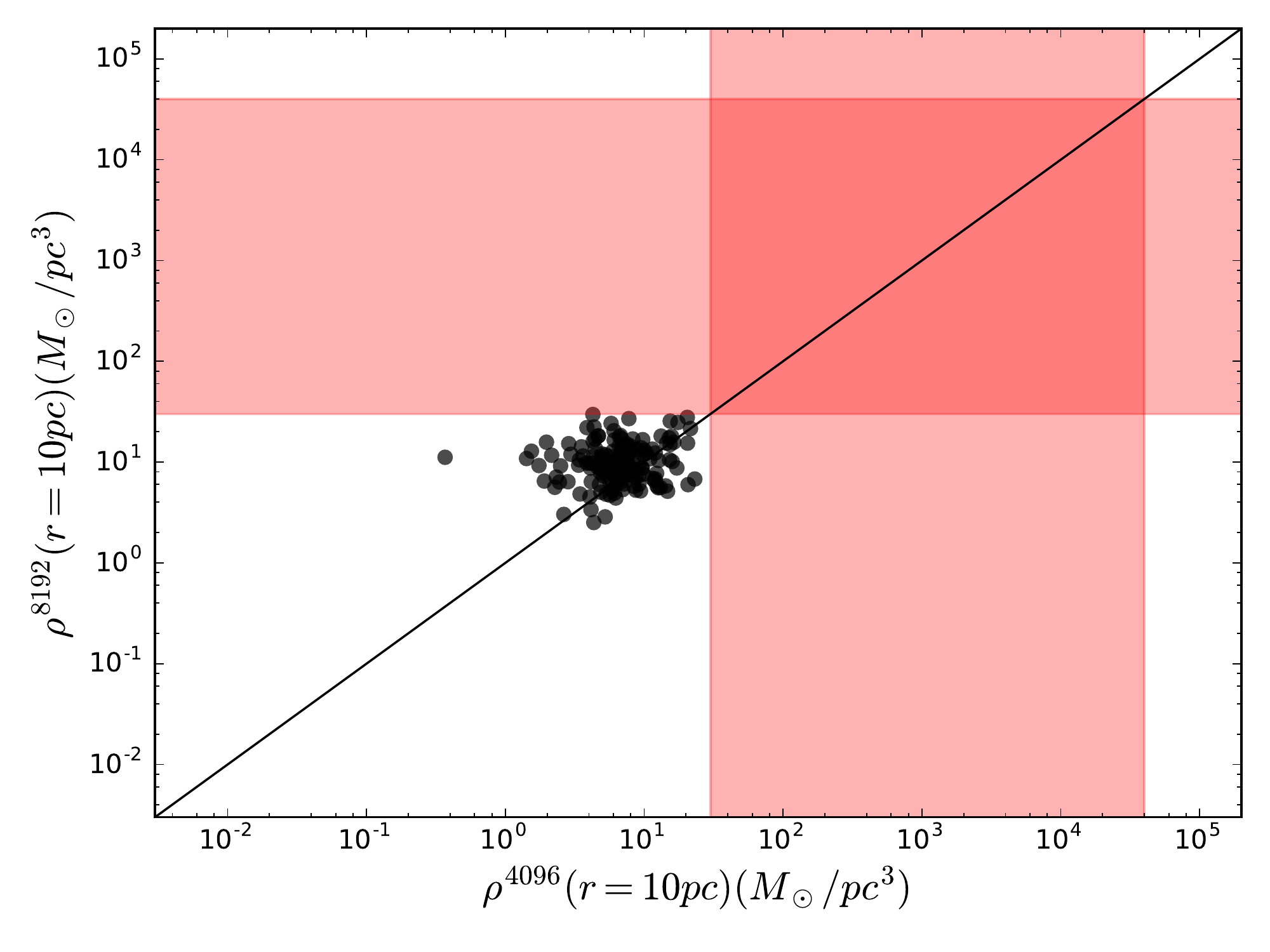}
\includegraphics[width=5.8cm]{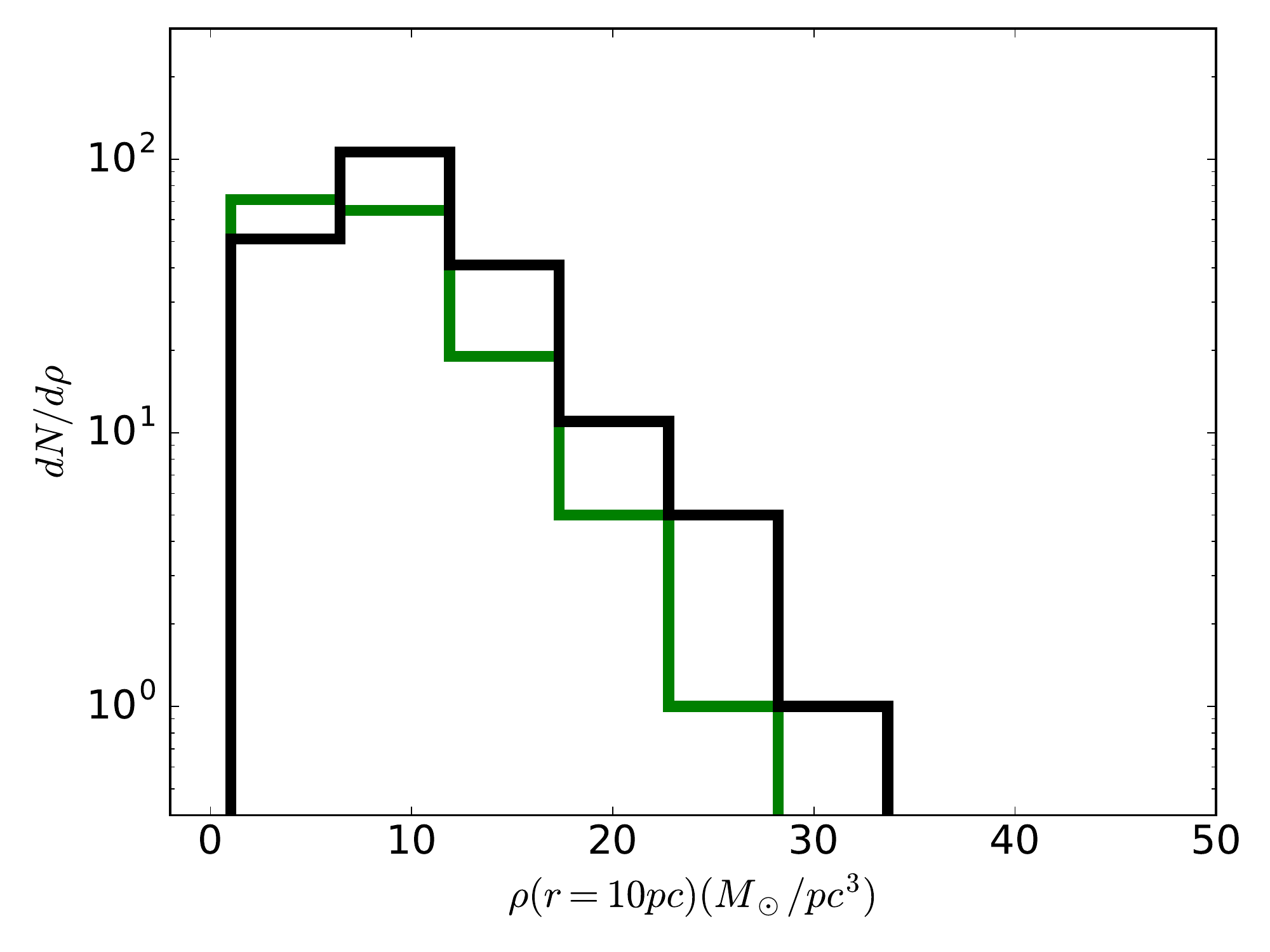}
\caption{{\it{Left}}: The extrapolated Einasto densities for the DSC candidates at $z=0$ for the 100M\_8192 versus 100M\_4096 (black points). The distribution of the derived densities for the DSCs are shown as the red shaded area. {\it{Right}}: Histogram of the distributions of the $z=0$ densities at extrapolated to 10 pc using the Einasto profile for 100M\_4096 (green) and 100M\_8192 (black).}
\label{FIG.Einastoscatter}
\end{figure*}

For the $z=0$ density profiles we use the Einasto profile \citep{Einasto:65,Navarroetal:04} with an assumed $\alpha = 0.18$ \citep{Springeletal:08}. For a chosen $\alpha$, the Einasto profile thus reduces to  a relatively simple function dependent on $r_{max}$ and $v_{max}$. We first choose the $N_{4096}$ subhalos in 100M\_4096 which have $M_{vir}(z_{peak}) > 10^9 M_\odot$ are resolved with $N>100$ at $z=0$. To check the convergence between the two simulations, we select the $N_{4096}$ most massive $z=0$ subhalos in 100M\_8192 with $M_{vir}(z_{peak}) > 10^9 M_\odot$ and compare the $v_{max}$, $r_{max}$ and $r_{vir}$ distributions for 100M\_4096 versus 100M\_8192 (Figure~\ref{FIG.convergence}). For all three parameters, we find good agreement between the distributions in 100M\_4096 and 100M\_8192. We next explicitly look at the densities of the extrapolated Einasto profiles at $10$~pc. In Figure~\ref{FIG.Einastoscatter} we show the distribution of the extrapolated $\rho(r=10 pc)$ for 100M\_4096 versus 100M\_8192 (left panel) and the distributions of $\rho(r=10 pc)$ for both runs (right panel). While there is not a one to one match between the halos, we find their distributions to be the same. In addition the density range inhabited by the DSCs (red shading) cannot be accounted for by the extrapolated dark matter densities in {\it{either}} simulation.

We therefore state the distributions of our $z=0$ Einasto profiles are convergent and any differences between individual halos does not change our result that the extremely high derived densities of the DSCs cannot be account for by dark matter alone.

\begin{figure*}
\centering
\includegraphics[width=8.9cm]{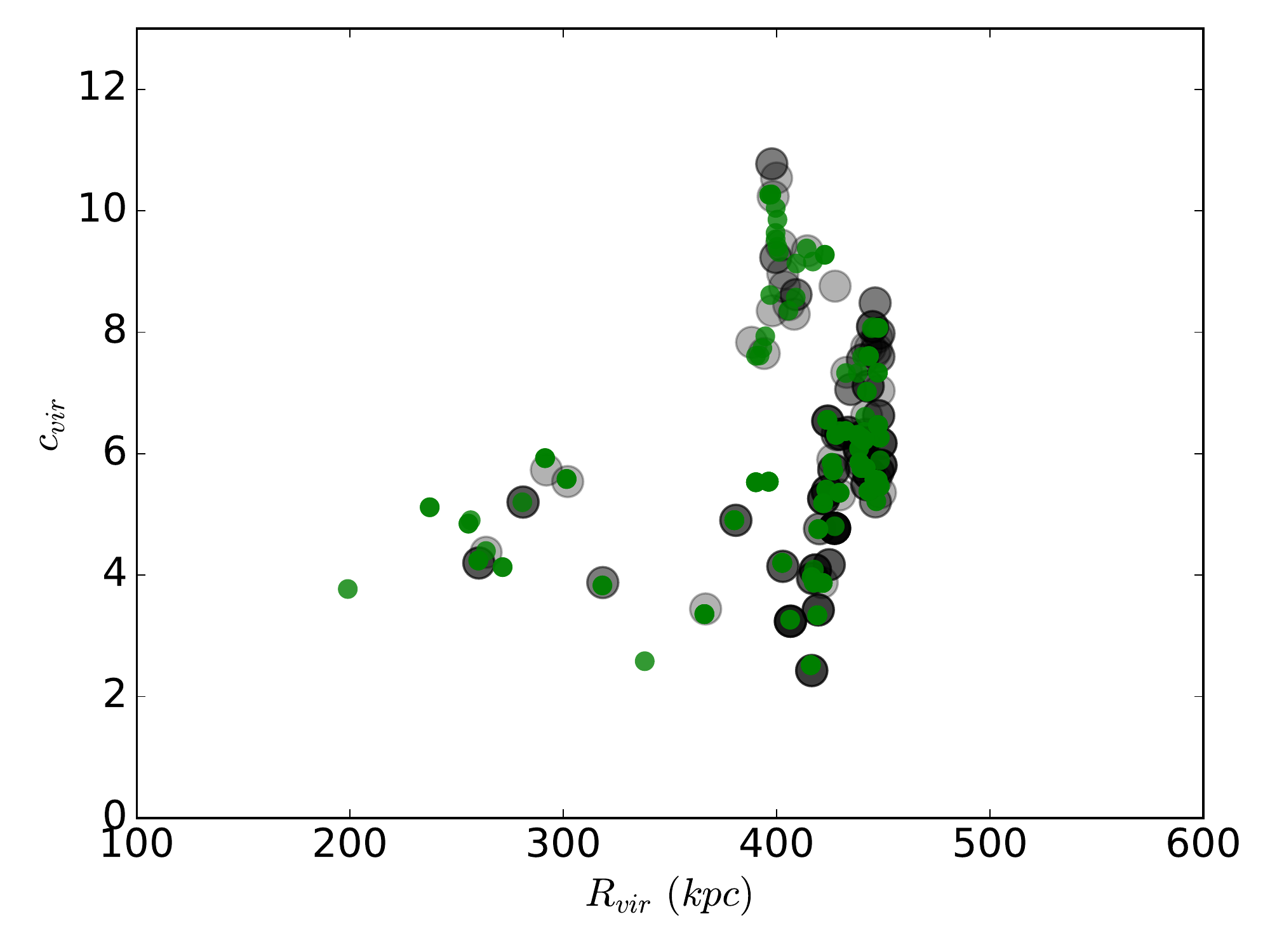}
\caption{Distribution of $c_{NFW}(z_{peak})$ versus $R_{vir}(z_{peak})$ for 100M\_4096 (green circles) and 100M\_8192 (black circles). }
\label{FIG.NFWconvergence}
\end{figure*}

\begin{figure*}
\centering
\includegraphics[width=5.8cm]{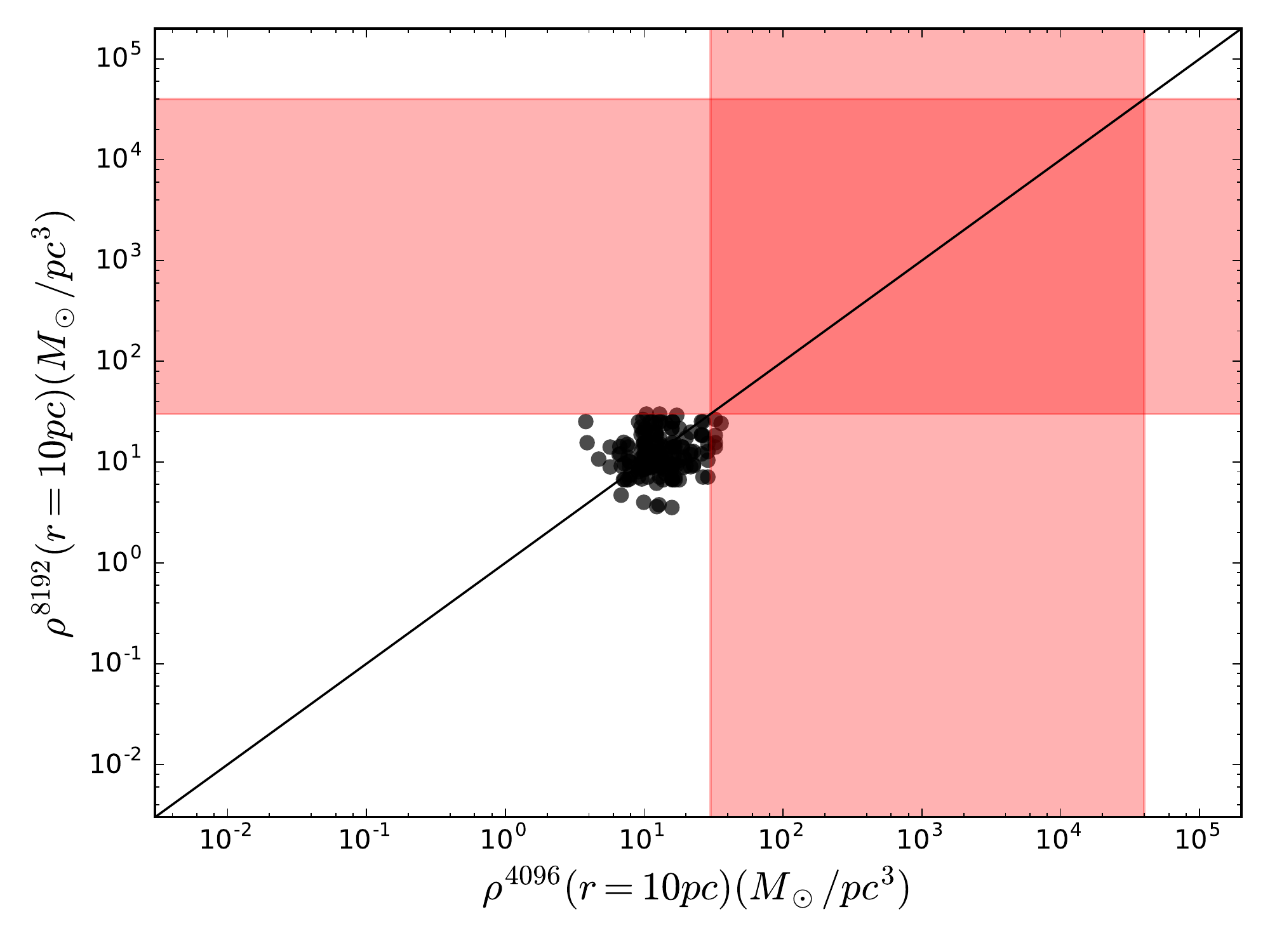}
\includegraphics[width=5.8cm]{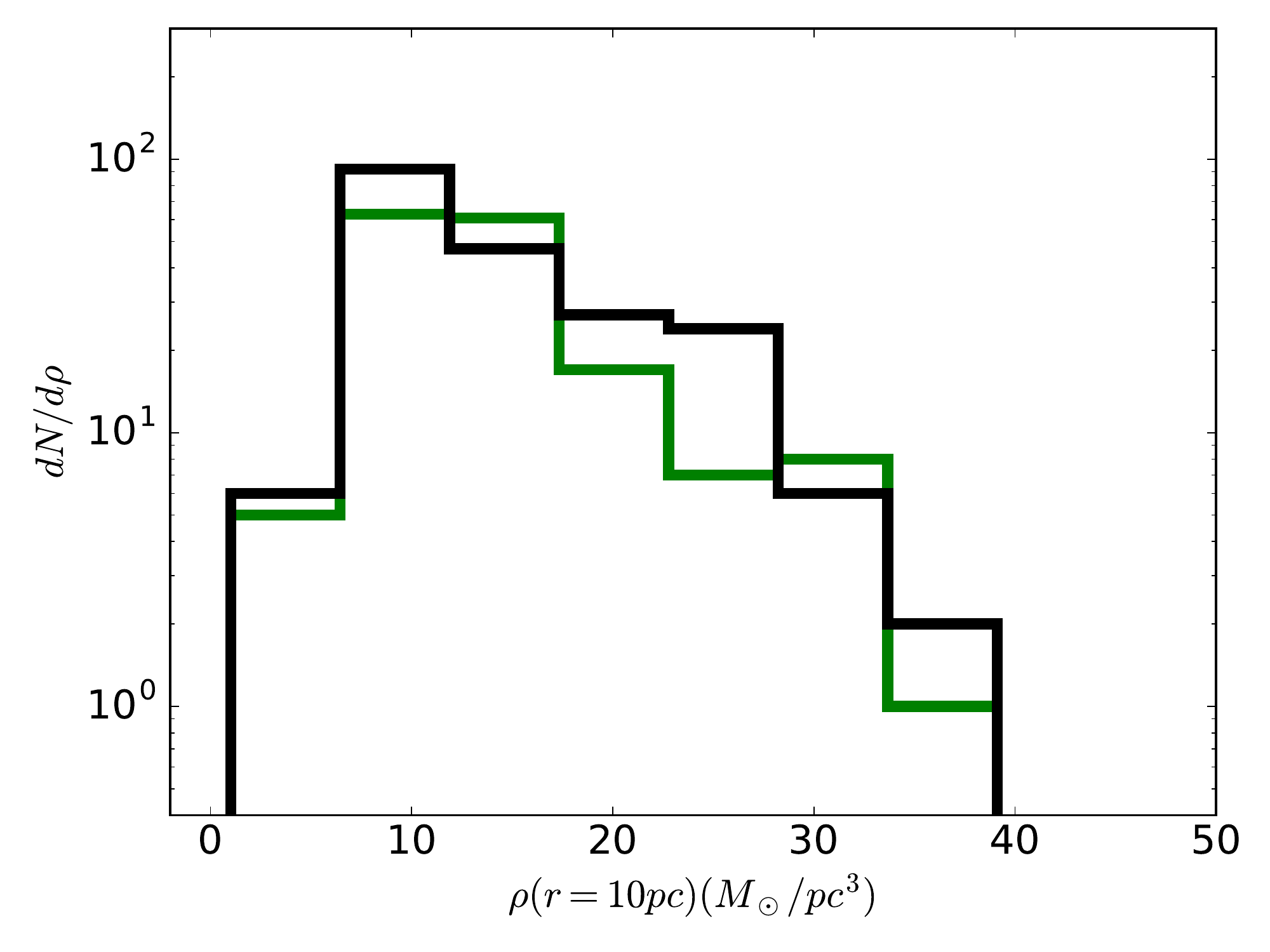}
\caption{{\it{Left}}: The extrapolated NFW densities for the DSC candidates at $z=z_{peak}$ for the 100M\_8192 versus 100M\_4096 (black points). The distribution of the derived densities for the DSCs are shown as the red shaded area. {\it{Right}}: Histogram of the distributions of the $z=z_{peak}$ densities at extrapolated to 10 pc using the NFW profile for 100M\_4096 (green) and 100M\_8192 (black).}
\label{FIG.NFWscatter}
\end{figure*}

We next determine the convergence of our NFW density profiles at $z_{peak}$. Since the NFW profile can be expressed as $\rho_{NFW}(r,c_{vir},R_{vir})$, we first confirm that the DSC candidate progenitors in 100M\_4096 and 100M\_8192 have the same distribution in $c_{vir}$ and $R_{vir}$ (see Figure~\ref{FIG.NFWconvergence}) before showing that they produce the same distribution of dark matter densities to $10$ pc (right panel of Fig~\ref{FIG.NFWscatter}). For both 100M\_4096 and 100M\_8192 the halos are well resolved at $z_{peak}$ with $N>1000$ and $N>10^4$ respectively.

We find the distribution of $c_{vir}$ versus $R_{vir}$ are extremely similar for 100M\_4096 and 100M\_8192. As with the Einasto profiles at $z=0$, we follow up the tests for convergence of the halo parameters with a confirmation that the distributions of the extrapolated densities at 10 pc are equivalent (see right panel of Figure~\ref{FIG.NFWscatter}. As with the $z=0$ Einasto profiles, we do not have a perfect one to one matching of our extrapolated densities, however the results from 100M\_4096 and 100M\_8192 are once again equilvalent. The dark matter densities at $z_{peak}$ were also at least an order of magnitude too low to account for the derived central densities of the DSCs.

As we show good agreement between 100M\_4096 and 100M\_8192 for both the $z=0$ Einasto profiles and the $z_{peak}$ NFW profiles in both their properties and extrapolated densities, we hold our extrapolation form the convergence radius of $600$ pc to within $20$ pc to be valid and robust for our analysis.

\label{lastpage}

\end{document}